\pdfoutput=1
\documentclass[usenatbib]{mn2e}
\usepackage{hyperref}
\usepackage{amsfonts,amsmath,amssymb}
\usepackage{txfonts}
\usepackage{graphicx}
\usepackage{bm}
\usepackage{ctable}
\usepackage{url}
\usepackage{breakurl}
\usepackage{fixltx2e} %% forces figure order to remain fixed (otherwise figure*'s can move out-of-order)
\usepackage{calc}
\newcommand\altaffilmark[1]{$^{#1}$}
\newcommand\altaffiltext[1]{$^{#1}$}
\newcommand\sh{Paper I}
\newcommand{\ws}{{\bf w}_{s}}
\newcommand{\wsh}{\hat{{\bf w}}_{s}}
\newcommand{\wsx}{{\rm w}_{s,x}}
\newcommand{\wsy}{{\rm w}_{s,y}}
\newcommand{\wsz}{{\rm w}_{s,z}}
\newcommand{\ts}{\tau_{s}}
\newcommand\ssomega{\boldsymbol{\mathsf{\Omega}}}

\newcommand{\acknowledgements}[1]{\begin{small}\section*{Acknowledgments}\end{small}{\noindent #1}\vspace{10pt}}
\newcommand{\revchng}[1]{{#1}}
\newlength\myheight
\newlength\mydepth
\settototalheight\myheight{Xygp}
\newcommand*\textimg[3]{%
  \settototalheight\myheight{Xygp}%
  \settodepth\mydepth{Xygp}%
 \raisebox{-#3\mydepth}{\includegraphics[height=#2\myheight]{#1}}
}
\newcommand{\figfold}{}
\newcommand{\specialcell}[2][c]{\begin{tabular}[#1]{@{}c@{}}#2\end{tabular}}

\title[]{Physical models of streaming instabilities in protoplanetary disks\vspace{-0.5cm}}

\author[Squire et al.]{
\parbox[t]{\textwidth}{ 
	Jonathan Squire\altaffilmark{1} \&\ 
	Philip F. Hopkins\altaffilmark{2} 
} 
\vspace*{6pt} \\
\altaffiltext{1}{Physics Department, University of Otago, Dunedin 9016, New Zealand} \\
\altaffiltext{2}{TAPIR, Mailcode 350-17, California Institute of Technology, Pasadena, CA 91125, USA\vspace{-0.3cm}}
}

\date{Submitted to MNRAS, ?, 2017\vspace{-0.6cm}}

\hypersetup{draft}
\begin{document}
\maketitle

\begin{abstract}
We develop simple, physically motivated models for drag-induced  dust-gas streaming instabilities, which are thought to be crucial for clumping grains to form planetesimals in protoplanetary disks. The models explain, based on the physics of  gaseous epicyclic motion and dust-gas drag  forces, the most important features of the streaming instability and its simple generalisation, the disk settling instability. Some of the key properties explained by our models include the sudden change in the growth rate of the streaming instability when the dust-to-gas-mass ratio surpasses one, the slow growth rate of the streaming instability compared to the settling instability for smaller grains, and the main physical processes underlying the growth of  the most unstable modes in different regimes. As well as providing helpful simplified pictures for understanding the operation of an interesting and  fundamental astrophysical fluid instability, our models may prove useful for analysing simulations and developing nonlinear theories of planetesimal growth in disks.
\end{abstract}

\begin{keywords}
\vspace{-1.0cm}
\end{keywords}

\vspace{-1.1cm}

%%%%%%%%%%%%%%%%%%%%%%%%%%%%%%
\section{Introduction}
%%%%%%%%%%%%%%%%%%%%%%%%%%%%%%%%%%
\begin{figure*}
\begin{center}
\includegraphics[width=1.0\textwidth]{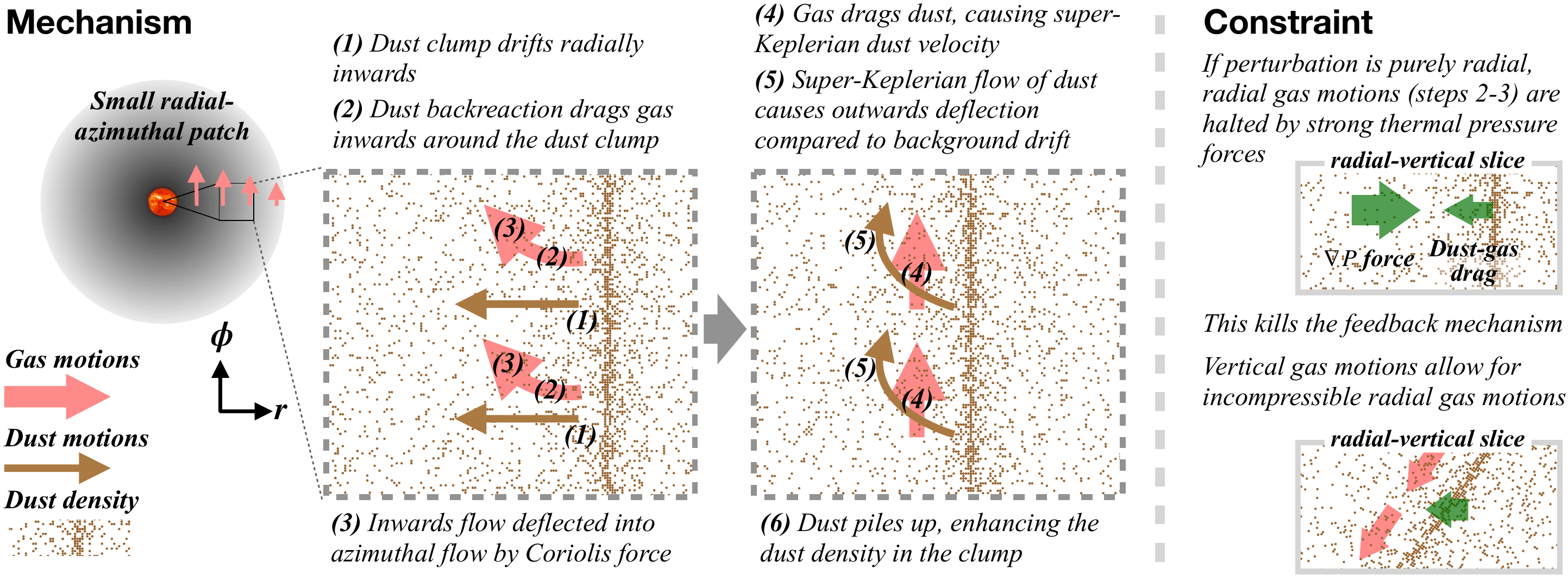}
\caption{Summary figure showing the key physical processes that enable the streaming instability. Steps (1)--(6) in ``Mechanism'' illustrate the azimuthal-radial plane to show how  
axisymmetric dust and gas motions can enhance a dust density maximum as it drifts inwards due to a 
combination of drag forces (1,2, and 4) and Coriolis and shear forces (3 and 5). However, such motions are only possible if perturbations vary vertically, as well as radially, as illustrated in ``Constraint.'' The sketched mechanism  is overly simplified in order to capture the key physical principles; figures \ref{fig:streaming.lowm}, \ref{fig:settling}, and \ref{fig:streaming.highm} give a more accurate representation of the  feedback loops that 
drive the instability across its different regimes.
  }
\label{fig:basic}
\end{center}
\end{figure*}
%%%%%%%%%%%%%%%%%%%%%%%%%%%%%%%%%%

Understanding the mechanisms that enable planetesimal formation within protoplanetary disks continues
to represent an outstanding  challenge in theoretical astrophysics \citep{Chiang2010}. 
An initial population of micron-sized dust grains entrained in the 
disk must coagulate and grow many orders of magnitude in size in order to collapse gravitationally. 
They must do so remarkably quickly, given the rapid infall of moderate-sized 
bodies \citep{Goldreich1973,Nakagawa1986} and the short lifetimes of observed disks \citep[see, e.g.,][]{Ansdell2017}.
A promising solution involves the clumping of grains through  collective  fluid-dynamical instabilities 
driven by the relative motion of dust through the gas \citep{Goodman2000}. The so-called ``streaming instability,'' 
discovered by \citet{Youdin2005}, is seen in nonlinear simulations to cause
significant clumping of dust, allowing gravitational collapse into planetesimals under the 
right conditions (see, e.g., \citealt{Johansen2007,Johansen2014,Simon2016} and references therein).
While there remain some potential issues---in particular, the super-solar metallicities and relatively 
large grain sizes that seem to be needed for robust planetesimal formation (e.g., \citealt{Johansen2009,Bai2010,Yang2017})---the general mechanism seems robust, particularly if accompanied by other grain concentration mechanisms such as
other collective dust-gas instabilities \citep{Squire2018a}, gas evaporation \citep{Williams2011}, or gas inhomogeneities \citep[and references therein]{Birnstiel2016}.

However, despite its likely astrophysical  importance, the streaming instability lacks  any simple, intuitive model 
for its operation, a situation that is somewhat unique among the broad  landscape of astrophysical  fluid and plasma instabilities (e.g., 
the magnetorotational instability; \citealp{Balbus1992a}).
In this article, we work to provide such models to explain  the linear growth  of dust-gas streaming instabilities, 
building on \citet{Squire2018a} (hereafter \sh) and a variety of previous studies by other authors \citep{Youdin2005,YoudinA2007,Jacquet2011,Auffinger2018,Jaupart2020}.
  Our models explain 
all the key features of the instabilities, including their dependence on physical parameters, their linear mode structure, 
and their growth rates.  As well as the traditional streaming instability of \citet{Youdin2005},
% and because it may also be important for planetesimal formation, 
we  also consider the related ``settling instability'' of \sh, which relies on very similar physics and
may clump small grains more efficiently than the streaming instability as they settle into the disk midplane.

Our primary motivation for this work is to  improve the general theoretical  understanding of a commonly studied mechanism 
that appears to be crucial to the development of planetary systems. For example, 
our models explain, based  on simple physical principles, the sudden change 
in the growth rate of the streaming instability as the dust-mass fraction surpasses one, and the
striking difference in growth rates between the streaming and settling instabilities.
In addition, such models may   
be able play a more practical role in future work by, for example, helping to diagnose and understand
grain clumping in more realistic nonlinear simulations. They may also be useful for developing nonlinear models of  
streaming-instability-generated turbulence, in a similar way to a variety of other astrophysical fluid instabilities, for example 
convection  \citep[e.g.,][]{Collins1989}, plasma instabilities \citep[e.g.,][]{Melrose1980}, or 
related dust instabilities \citep{Moseley2019,Seligman2019,Hopkins2020}.

Before discussing the streaming instability in detail, we feel it helpful to provide a single, simplified sketch of its mechanism
 in figure \ref{fig:basic}. This illustrates how the basic physical ingredients of drag and Coriolis forces can conspire to enhance  
 a drifting dust density clump, so long as the clump also has vertical structure. We believe that this is 
the simplest reasonable representation of the physics of the streaming instability (although it remains overly simplified), and
may satisfy the reader who desires a basic understanding of its key features. 
The physics driving various unstable modes in different regimes is thoroughly described throughout the rest of the paper.

\subsection{The philosophy of this article}

Given our desire to formulate simple, physical explanations  for dust clumping in disks, the  approach 
of this article  is to avoid, wherever possible, detailed mathematical presentation of growth rates
and mode structure. Rather, we attempt  to distill each case down to the simplest possible explanation that 
correctly captures its physical characteristics. Our models are thus deliberately rather qualitative. Nonetheless, 
we wish to emphasise that each was derived using the linearised equations
of motion for the coupled dust-gas system, employing   understanding  gained using 
the ``resonant drag instability'' (RDI) theory of \citet{Squire2018}. As an example, the response of the dust density to an epicycle, which
 will be 
introduced in  \S\ref{sub: gas on dust} and used extensively throughout the article, can be derived  through the direct solution 
of the four linearised dust equations, forced by a periodic gas velocity.  
We invite the reader interested in further mathematical details to consult \sh. A related 
mathematical discussion, which focuses primarily on the settling instability, is found in \citet{Zhuravlev2019}, 
and the recent papers of \citet{Pan2020,Pan2020a} discuss a variety of similar concepts from a somewhat more mathematical perspective.  

\revchng{With the same spirit, throughout the article we will neglect all non-ideal physical effects that
would complicate the discussion; we consider only a single population of grain sizes and neglect 
the effects of  background gas turbulence, gas viscosity, and dust diffusion or dispersion. 
Such effects can be very important, and their neglect here is motivated only by our desire to keep 
the models as simple as possible. Indeed, a distribution of grains has been shown to strongly 
impede the growth of the streaming instability in the low-dust-to-gas-ratio regime  \citep{Krapp2019}, while
turbulence and/or viscosity can significantly limit the applicability of the instability in many scenarios \citep[e.g.,][]{Squire2018a,Umurhan2019,Chen2020a,Jaupart2020,Zhuravlev2020,Krapp2020}. Our models could nonetheless prove useful as the basis for more 
complex models of instabilities in systems with distributions of grain sizes, while
turbulence, viscosity, or dust diffusion could likely be straightforwardly added  if desired 
(for example,
viscosity will impede the growth of  modes below some scale).}

Following a brief description of the local approximation, dust drifts (\S\ref{sub: conventions}), and
the physics of coupled dust-gas dynamics (\S\ref{sub: equations}), our 
presentation starts with an analysis of how dust responds to gas epicycles (\S\ref{sub: gas on dust}), and how 
the gas responds to dust density perturbations (\S\ref{sub: dust on gas}). Coupled to the idea
of resonance between drifting dust and gas modes (the basis for RDI theory),
 these responses explain the characteristics of the low-dust-to-gas-ratio streaming instability in \S\ref{sec: streaming} and the 
  settling  instability in \S\ref{sec: settling},  clarifying why they 
exhibit vastly different growth rates for small grains. 
As  noted in \citet{Youdin2005} and \sh, 
the character of the streaming instability changes substantially when the dust-mass fraction is 
larger than unity, and this regime must be considered separately. We explain this transition and
the characteristics of the high-dust-to-gas-ratio streaming instability in \S\ref{sec: high mu streaming}. 
Our model explains its large growth rate, which is likely fundamental to the planetesimal formation process. 
Our models for each case are sketched in figures~\ref{fig:streaming.lowm}, \ref{fig:settling}, and \ref{fig:streaming.highm}, 
which are intended to be understandable without detailed reference to the text.

\section{Dust-gas interaction in disks}\label{sec: basics}

\begin{table*}
\begin{center}
 \begin{tabular}{||c c c ||} 
 \hline
 Symbol &  Description and/or definition & Graphical representation and/or  notes\\ [0.5ex] 
 \hline\hline
     $r$, $\phi$, $z$  & Radial, azimuthal, vertical: global coordinates  &  \specialcell{All models neglect global radial and vertical variation of background.\\By rescaling  grain size and wavelength, they apply to any  location in the disk. } \\
    \hline
    $x$, $y$, $z$  & Radial, azimuthal, vertical: local coordinates  &Solid grey panel borders \textimg{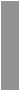}{1.6}{2}  show $x$-$z$ plane; dashed-grey borders \textimg{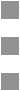}{1.6}{2} show $x$-$y$ plane  \\
    \hline
      $\mu$  &    Dust-to-gas continuum mass density ratio $\rho_{d}/\rho$ & The streaming instability is fundamentally different for $\mu<1$ and $\mu>1$ \\ 
              \hline
%   $t_{s}$  &   Dust stopping time (drag time) & The drag regime (Stokes versus Epstein) is unimportant  \\ 
% \hline
  $\tau_{s}$  &   Stokes number $\tau_{s} = \Omega\,t_{s}$ (stopping time in local  units)  & $\tau_{s}\lesssim 1$ grains interact strongly with the gas.  \\ 
\hline
 $\Omega$  &   Keplerian rotation frequency & Dust interacts with gas epicycles  ($\omega=\pm \hat{k}_{z}\,\Omega$) to cause instability. \\
    \hline
  $\eta$  &   Disk pressure support, $\eta\!\sim (h/r)^{2}$ for disk scale height $h$ & Instabilities do not depend explicitly on $\eta$ except through $\wsz/\wsx$ (\S\ref{subsub: eta definitions})\\ 
    \hline\hline
    $\bm{k}$  & Wavenumber of (axisymmetric) mode  $\bm{k}=(k_{x},0,k_{z})$ &  Direction ($\hat{\bm{k}}$) shown with a grey arrow \textimg{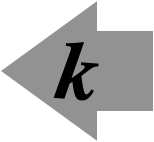}{2}{3}\\ 
 \hline
    $\omega$  & Frequency of mode ($\Im(\omega)>0$ implies instability) &  Bottom panels show time evolution of mode at position marked with \textimg{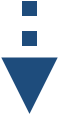}{2}{3}. \\
 \hline
   $\rho$  &    Gas density & Models assume incompressibility due to strong gas pressure forces (\S\ref{sub: equations}) \\ 
    \hline
$\rho_{d}$  &    Dust continuum mass density, $\delta \rho_{d}$ is its perturbation  & Relative density illustrated with brown dots \textimg{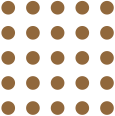}{2}{3} \\ 
\hline
  $\bm{u}$ &    Perturbed gas velocity ($\bm{u}=\bm{U}-Sx\,\hat{\bm{y}}-\Delta\bm{U}_{0}$)  & \specialcell{The in-plane flow lines of $u_{x}$  and $u_{z}$ are shown with black arrows \textimg{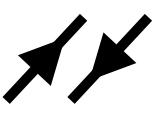}{1.7}{2.4}. \\ Regions with $u_{y}>0$ ($u_{y}<0$) are colored as \textimg{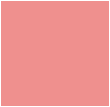}{1.3}{1.5} (\textimg{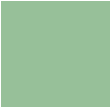}{1.35}{1.5}).}\\ 
  \hline
  $\bm{v}$ &    Perturbed dust velocity ($\bm{v}=\bm{V}-Sx\,\hat{\bm{y}}-\Delta\bm{V}_{0}$)  & Deflection of $v_{y}$ into $v_{x}$ due to Coriolis force shown with \textimg{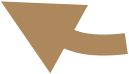}{1.5}{2}. \\ 
   \hline
   $\ws$  & Equilibrium dust-gas drift (see \S\ref{subsub: eta definitions}) & \specialcell{$\ws\approx \wsx \hat{\bm{x}}$ for streaming instability; $\wsz\neq0$ for settling instability. \\Shown with \textimg{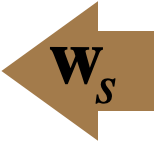}{2}{3}.}\\ 
   \hline
   $\bm{F}$ & Backreaction force from dust onto gas &  $\bm{F}\propto\delta \rho_{d}\ws$, caused by $\delta \rho_{d}$ because of dust drift. Shown with \textimg{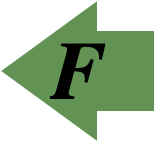}{2}{3}.\\
     \hline
 \end{tabular}
\end{center}
\caption{Important symbols and definitions used throughout this article. The right-hand column notes useful considerations relating to the symbol in question and/or   shows how  symbols  are illustrated graphically in  figures~\ref{fig:basic.gas2dust}--\ref{fig:streaming.highm}.  The key background parameters that control  instability properties are $\mu$ and $\tau_{s}$.}
\label{tab:}
\end{table*}

\subsection{Conventions and definitions}\label{sub: conventions}

The instabilities that concern us in this article are all local in character. This means that rather 
than having to describe specific  models for  global density, temperature, and metallicity 
variations in a disk, we can consider just a local 
patch of near-Keplerian disk with constant background gas and dust densities
and a constant linear shear profile. 
Of course, we assume that all instabilities studied  vary on small scales compared to the
global disk variation, and if this is found \emph{a-posteriori} to not be true, our analysis is
not valid. The great advantage of the local approach is simplicity, in particular, that our 
system depends only on two parameters: the dust size and the dust-to-gas mass ratio.
%(if we allow wavelengths to be rescaled with changes in background pressure gradient; see below). 
%For example,  similar 
%parameters will be applicable either in high-density regions close to the star with larger grains, 
%or lower-density, far out regions with smaller grains. 
\revchng{We further assume that perturbations to the local equilibrium are very small, allowing one to derive linear 
equations for the spatiotemporal evolution of any quantity (velocity, density, etc.). Solutions to these equations can then be completely characterised by assuming that the perturbation of each fluid quantity follows the 
spatiotemporal variation ${f}_{\rm spat}(\bm{x},t)=f(\bm{k},\omega) \exp(i \bm{k}\cdot\bm{x}-i\omega t) + \mathrm{conj.}$, where $
{f}_{\rm spat}(\bm{x},t)$ is a fluid perturbation (e.g., perturbed gas velocity or the variation in dust density), and $f(\bm{k},\omega)$ is
its complex Fourier amplitude. Due to linearity, the Fourier amplitudes drop out and $\omega$  satisfies a simple polynomial equation (the dispersion relation) 
with the 
wavevector $\bm{k}$ (magnitude $k$ and direction $\hat{\bm{k}}$) as a parameter. Further, since we are free to study the stability of any chosen mode separately from all other modes, the spatial variation of all quantities in our models is naturally sinusoidal.}
In keeping with convention, we use the directions $\hat{\bm{x}}$, $\hat{\bm{y}}$, and $\hat{\bm{z}}$ to 
refer to the radial, azimuthal, and disk-normal (vertical) directions in the local frame, while $(r,\phi,z)$ refer to 
the global coordinate system. A number of useful definitions, including illustrations of the symbols used in later 
figures, are given in table \ref{tab:}. 

\subsubsection{Dust properties}\label{subsub: dust properties}

We assume the dust to be a pressureless fluid with average continuum mass density $\rho_{d}=\mu \rho$, where $\rho$ is the 
gas density and $\mu$ is the dust-to-gas  ratio. 
The dust interacts with the gas through drag forces governed by the stopping time $t_{s}$, which is the approximate time 
it takes a particle to come to rest in the frame of the gas. As in previous works it is helpful to 
parameterise the dust size with the \emph{Stokes number}, $\tau_{s} \equiv t_{s}\Omega$, where $\Omega$ is the Keplerian 
frequency in our local patch of disk. Particles with $\tau_{s}\lesssim 1$ can be considered ``strongly coupled'' to 
the gas (drag forces are larger than rotational forces) and are the focus of this article. Understanding the coagulation of those in the 
range $10^{-4}\lesssim \tau_{s}\lesssim 1$, with
 physical sizes from the submillimeter scale up to somewhat less than a meter depending on the location in the disk, presents a number of interesting  challenges to planet formation theory \citep{Chiang2010,Carrera2015,Yang2017}.
It transpires that the detailed functional form of the dependence of $\ts$ on ambient gas properties---for example, whether the grains obey the Epstein or Stokes drag law---is
 not important for the discussion here, and  our results can apply equally well at any physical location in the disk, so 
long as the dust size is scaled appropriately  (the 
same $\tau_{s}$ corresponds to smaller particles further out in the disk). 
In formulating  our models, we will sometimes assume $\ts\ll1$, 
which is useful for understanding the general motions and  forces on the dust and gas. However, the basic
mechanisms we discuss for the streaming and settling instabilities are qualitatively applicable for 
all grains with $\ts\lesssim 1$.

%\pfh{You don't use the usual $\eta$ parameter here. Is $h/r$ actually exactly identical? I don't think so, just dimensionally. Since there is so much 
%literature using the parameter $\eta$ you should probably introduce it even if only to twiddle it away, or explain how it in its usual form relates to $h/r$. 
%Also worth reiterating that we are assuming things like thermal pressure dominates vertical/pressure of disk (weak turbulence, B-fields) and the disk 
%is a Keplerian disk (though changing the potential has no important effect on the results) in near-equilibrium with quasi-circular orbits (expanding about circular to leading order), with the gas in vertical equilibrium (so $h/r = c_{s}/v_{k}$), etc. these can all be stated pretty briefly.}

\subsubsection{Pressure gradient and dust drifts}\label{subsub: eta definitions}

The fundamental driver of the streaming and settling instabilities is the presence of global 
gas pressure gradients $\partial P/\partial \ln r$ and $\partial P/\partial \ln z$. The radial gradient ($\partial P/\partial \ln r$) causes the gas motions to differ slightly \revchng{(by $\Delta\bm{U}_{0}$)} from circular Keplerian orbits (velocity $U_{K}$), with the proportional difference in azimuthal velocity ($\Delta U_{0,y}/U_{K}$),
\begin{equation}
\eta \equiv- \frac{\Delta U_{0,y}}{U_{K}}\approx - \frac{\partial P/\partial \ln r}{2\rho U_{K}^{2}} \sim \frac{c_{s}^{2}}{U_{K}^{2}}\sim \frac{h^{2}}{r^{2}},\label{eq:eta.def}
\end{equation}
 where $c_{s}$ is the gas sound speed and  $h/r$ is the disk aspect ratio. Equation~\eqref{eq:eta.def} has assumed a thin disk in vertical equilibrium, with
 pressure support dominated by thermal forces (weak turbulence and magnetic fields). 
\revchng{ The pressure-support-induced velocity difference $\Delta\bm{U}_{0}$, coupled to the
gas-dust drag forces, 
induces an equilibrium velocity of the dust compared to Keplerian orbits, denoted $\Delta \bm{V}_{0}$. Because $\Delta\bm{V}_{0}\neq \Delta\bm{U}_{0}$, this causes a bulk drift of the dust in the frame of the gas, $\ws\equiv \Delta\bm{V}_{0}- \Delta\bm{U}_{0}$, which provides the free-energy source for instability.} In the midplane of the disk, where $\partial P/\partial \ln z=0$, this is  \citep{Nakagawa1986},
\begin{equation}
\frac{\ws}{\eta U_{K} r} \approx -2 \frac{(1+\mu)\ts}{(1+\mu)^{2}+\ts^{2}} \hat{\bm{x}} + \frac{\ts^{2}}{(1+\mu)^{2}+\ts^{2}}\hat{\bm{y}},
\label{eq:NSH.xy}
\end{equation}
which is dominated by inwards radial drift for smaller particles; $\ws/(\eta U_{K}) \approx -2\ts\hat{\bm{x}}$ for
$\ts\ll1$, $\mu\ll1$. If the dust is separated from the disk midplane, it also drifts towards the midplane due to vertical pressure gradients  $\partial P/\partial \ln z$, with velocity 
 \begin{equation}
\frac{\wsz}{\eta U_{K} r} \approx \eta^{-1/2} \frac{\ts}{1+\ts}, \label{eq:NSH.z}
\end{equation}
which is significantly larger than the radial drift for small particles ($\eta^{-1/2}\sim r/h\gg1$) owing to the larger effective vertical gravitational 
force. Although  $\ws$ depends explicitly  on the gas pressure gradient (through $\eta$),
by using $\eta r$ as a length unit and $\Omega$ as a time unit, this dependence drops out, 
except in the relative size of $\wsz$ compared to $\wsx$. Equivalently, the dependence of a linear mode
 on $\eta$  can be captured by simply rescaling its wavelength by $\eta r$ and  timescales by $\Omega$.
We  can thus  ignore 
the dependence of our results on $\eta$ (equivalently $\partial P/\partial \ln r$ or $h/r$) throughout our analysis, so long as we assume that it is relatively small.
Note that the local approximation must break down for modes with longer wavelengths  than the vertical scale height, $k=|\bm{k}| \lesssim \eta^{1/2}(\eta r)^{-1}$.

\subsection{Equations of motion for the gas and dust}\label{sub: equations}

Although we will not directly solve the fluid equations 
in this article, it is helpful to present them here in order to highlight the most important terms and their effects. 
As discussed above,  we consider a local patch of disk in a frame moving
with the local Keplerian velocity. We also assume the gas to be {locally} incompressible, which  is a good approximation 
for all modes of interest (see \S\ref{subsub: incompressibility} below).  
%It remains unclear whether instabilities of the \citet{Goodman2000} type can operate in realistic settings, and what the necessary ingredients are (e.g., a turbulent dust layer; see also \citealt{Chiang2010}).
The gas equations are then,
\begin{gather}
\nabla \cdot \bm{U}=0,\label{eq: gas div=0}\\
\frac{\partial}{\partial t}\bm{U} + \bm{U}\cdot \nabla\bm{U} +2\Omega \,\hat{\bm{z}}\times \bm{U} = -S U_{x}\,\hat{\bm{y}} -\frac{\nabla P}{\rho}-\frac{\rho_{d}}{\rho}\frac{\bm{U}-\bm{V}}{t_{s}} ,\label{eq: gas NL}
\end{gather}
while the dust satisfies, 
\begin{gather}
\frac{\partial}{\partial t}{\rho_{d}} + \nabla\cdot (\rho_{d} \bm{V} )=0,\label{eq: dust density} \\
\frac{\partial}{\partial t}\bm{V} + \bm{V}\cdot \nabla\bm{V} +2\Omega \, \hat{\bm{z}}\times \bm{V} = -S V_{x}\, \hat{\bm{y}} -\frac{\bm{V}-\bm{U}}{t_{s}}.\label{eq: dust NL}
\end{gather}
\revchng{Here  $\bm{U}$ and $\bm{V}$ are the local gas and dust velocities, with $S=-(3/2)\Omega$ the local Keplerian velocity shear. The velocities are 
decomposed as $\bm{U}=Sx\,\hat{\bm{y}} + \Delta \bm{U}_{0}+\bm{u}$ and $\bm{V}=Sx\,\hat{\bm{y}} + \Delta \bm{V}_{0}+\bm{v}$; i.e., each has a shear contribution $Sx\,\hat{\bm{y}}$, an equilibrium drift contribution $\Delta \bm{U}_{0}$ (or $\Delta \bm{V}_{0}$) due to the background pressure gradient (see \S\ref{subsub: eta definitions}), and a perturbed contribution $\bm{u}$ (or $\bm{v}$). It is the perturbed velocities $\bm{u}$ or $\bm{v}$ that will be considered in our analyses below. The gas density density is assumed constant and denoted by $\rho$, while the gas pressure $P$ contains both a perturbed part that is chosen to enforce $\nabla\cdot\bm{u}=0$, and the equilibrium contribution discussed in \S\ref{subsub: eta definitions} (this is balanced by $\Delta \bm{U}_{0}$ and $\Delta \bm{V}_{0}$ related terms).  The
dust density is $\rho_{d}$, while its perturbation is denoted $\delta \rho_{d}=\rho_{d}- \rho_{d0}$ (where $\rho_{d0}$ is the mean dust density $\mu=\rho_{d0}/\rho$).}

The physical effects that will turn out to be most relevant to the ensuing discussion are: 
\begin{description}
\item[\textbf{(i) Dust advection}] \revchng{In the equilibrium frame of the gas (velocity $Sx\,\hat{\bm{y}}+\Delta\bm{U}_{0}$), the linearised dust continuity equation \eqref{eq: dust density} becomes
\begin{equation}
\left(\frac{\partial}{\partial t} + \ws \cdot \nabla\right)\frac{\delta \rho_{d} }{\rho_{d0}} + \nabla\cdot\bm{v} = 0,
\end{equation}
showing that 
the offset in mean velocities between the gas and the dust (equations~\eqref{eq:NSH.xy} and \eqref{eq:NSH.z}) advects  dust density perturbations with the velocity $\ws$.}
\item[\textbf{(ii) Pressure forces}] Any attempt to create motions that compress the gas create a large pressure 
force that opposes these motions ($-\nabla P$). This causes gas motions to be only  weakly compressible on the scales of interest, justifying  our assumption of incompressibility (see \S\ref{subsub: incompressibility}). 
This is not the case for dust motions.
\item[\textbf{(iii) Coriolis forces}] Both the gas and the dust experience \revchng{Coriolis and velocity shear} forces due to the transformation to the rotating,  shearing frame. This generates radial from azimuthal velocities, and vice versa, of the form 
\begin{gather}
\frac{\partial}{\partial t} v_{x} = 2 \Omega v_{y}+\dots,\nonumber\\
\frac{\partial}{\partial t} v_{y} = -(2\Omega+S) v_{x}+\dots=-\frac{1}{2}\Omega v_{x}+\dots \label{eq: coriolis}
\end{gather}
(likewise  for $\bm{u}$). \revchng{The key effect that enables dust compressions is the generation of radial from azimuthal velocities, which 
relies only on the Coriolis force.}\footnote{\revchng{In fact, all of the instabilities we consider here continue to operate in a very similar 
way in the absence of velocity shear, albeit with modified frequencies and growth rates.}}
\item[\textbf{(iv) {Strong} drag forces}]We focus on the $\ts \ll 1$ ($t_{s}\Omega \ll1$) regime, where drag forces are strong 
compared to  the $\sim\!\!\Omega^{-1}$ timescales of motions. This means that dust rapidly reaches its ``terminal velocity'' \citep{Youdin2005,Laibe2014} where the relative drift velocity of dust compared to gas is  determined by the pressure gradient. 
\item[\textbf{(v) Dust density force {(backreaction)}}] The final term in Eq.~\eqref{eq: gas NL} is the backreaction of the dust on the gas motions. Because of the mean dust drift $\Delta\bm{V}_{0}-\Delta\bm{U}_{0}=\ws$, a local increase or decrease in 
dust density generates an effective force per unit mass $\bm{F}$ on the gas (compared to its equilibrium),  \begin{equation}
\bm{F}=(\rho_{d}-\rho_{d0}) \frac{\ws}{\rho t_{s}}=\mu\,\frac{\delta \rho_{d} }{\rho_{d0}} \frac{\ws}{\tau_{s}}\Omega,\label{eq: F.def}.
\end{equation}
\end{description}

\revchng{
\subsubsection{Justifying the assumption of incompressibility}\label{subsub: incompressibility}
A perturbation of wavenumber $k$, with a gas density variation $\delta \rho/\rho$, will exert 
a force per unit mass on the surrounding gas of order $\sim\! k c_{s}^{2}(\delta\rho/\rho)$. Similarly, the 
force per unit mass 
exerted due to a drifting dust-density perturbation (the dust-feedback mechanism that
enables streaming instabilities; see \S\ref{sub: dust on gas}) is $\sim \mu( \ws/\tau_{s})(\delta \rho_{d}/\rho_{d0})\Omega\sim\max(\mu,1)\eta^{1/2}h^{-1}c_{s}^{2} (\delta \rho_{d}/\rho_{d0})$. Here, we have used the scalings from Eqs.~\eqref{eq:eta.def}--\eqref{eq:NSH.xy}, assuming in-plane radial drift $\wsx/(\eta U_{K}r) \approx -2\tau_{s}/\max(\mu,1)$. 
Comparison of the two estimates shows that  a gas density perturbation 
$\delta\rho/\rho\sim\max(\mu,1)\eta^{1/2}(kh)^{-1} (\delta \rho_{d}/\rho_{d0})$ will affect the gas  with 
similar force to the dust perturbation $\delta \rho_{d}/\rho_{d0}$. Since $\max(\mu,1)\eta^{1/2}(kh)^{-1} \ll1$, we see that only very small gas density perturbations 
are needed to create pressure forces similar to the dust backreaction force that drives the instability, showing that
the gas can be very well approximated as  incompressible.\footnote{In the settling instability, where $\wsz\sim \eta^{-1/2}\wsx$, 
the factor becomes $\sim\max(\mu,1)(kh)^{-1}$, but this is still much smaller than unity for all modes considered.}
}

\revchng{
More
explicitly, both \sh\ and \citet{YoudinA2007} showed through numerical and analytic calculations of the dispersion relation that allowing or removing compressibility has a negligible effect on the behaviour of the streaming-instability. Note that the toy secular instability 
of \citet{Goodman2000} requires compressibility of the gas surface density, so is not described 
by our models.  
}

\section{Resonant forcing of linear systems}\label{sec: resonances}

In this section, we discuss heuristically how instabilities of the coupled dust-gas system  arise from 
the resonant driving of the dust density perturbations by the gas, coupled with resonant driving of gas modes by the dust. 
The discussion tells us that in order to understand the mechanisms for the streaming and 
settling instabilities, we should consider how these resonant driving mechanisms  feed back on each other. 
If the dust perturbation generated by a gas mode has the  effect of resonantly forcing this original mode, the coupled system 
will be linearly unstable.

\subsection{Secular growth at resonance}\label{sub: secular.growth}

It is well known that applying a time-varying force to an oscillator system leads to a particularly strong 
response when the frequency of the forcing matches that of a natural mode of the system. As a particularly 
simple example, the system 
\begin{equation}
\partial_{t} f(t) + i \omega f(t) = F e^{-i \omega_{F} t},\quad f(0)=0,\label{eq:forced system}
\end{equation}
has the (bounded) solution $f(t) = i F (\omega-\omega_{F})^{-1}(e^{-i\omega t}- e^{-i\omega_{F} t})$ for 
off-resonant ($\omega\neq \omega_{F}$) forcing, while resonant forcing ($\omega= \omega_{F}$)
leads to (unbounded) continuous secular growth, $f(t) = F t e^{-i\omega t}$.
In the more complex case where $f(t)\rightarrow{\bm f}(t)$ is a system of variables, $i\omega \rightarrow i\ssomega$ represents a  general linear operator,  and $F\rightarrow \bm{F}$ is a vector of forcing amplitudes,
the strength of the effect of forcing at resonance is governed by $\bm{\xi}^{L}\cdot {\bm F}$, where $\bm{\xi}_{L}$ is the left eigenvector 
of $i\ssomega$  that is associated with the frequency of the forcing ($\omega_{F}$). 
In other words, the system will execute unbounded secular growth $\bm{f}(t) \propto t e^{-i\omega t}$ so long 
as two conditions are met: (i) the forcing frequency $\omega_{F}$ matches an eigenfrequency $\omega$ of $i\ssomega$; and 
(ii) the left eigenmode (termed  $\bm{\xi}^{L}$) of $i\ssomega$ associated with  $\omega=\omega_{F}$  is not orthogonal to the forcing direction, $\bm{\xi}^{L}\cdot {\bm F}\neq0$.
Physically, the idea is that in order to resonantly drive a given eigenmode of the system, the applied forcing
must both match its frequency and  lie along a direction that is able to force the motions involved in the mode.\footnote{Note that
in the $1\times1$ system of Eq.~\eqref{eq:forced system}, $\xi^{L}=1$, so condition (ii) is always met if $F\neq0$.}

The resonant drag instability (RDI) theory of \citet{Squire2018} quantifies how such resonant interactions
lead to linear \emph{instabilities} of coupled linear systems. The relation 
of their result to the above arguments can be understood by considering a system such as 
\begin{gather}
\partial_{t} {\bm f}_{1} + i\ssomega_{1} \cdot {\bm f}_{1} = {\bm F}_{2\rightarrow 1},\nonumber \\ 
\partial_{t} {\bm f}_{2} + i\ssomega_{2} \cdot {\bm f}_{2} = {\bm F}_{1\rightarrow 2}, 
\end{gather}
where the homogenous part of system 1 ($\partial_{t} {\bm f}_{1} + i\ssomega_{1} \cdot {\bm f}_{1} =0$) represents the linearized equations
of motion of gas in the absence of dust, while the homogenous part of system 2 ($\partial_{t} {\bm f}_{2} + i\ssomega_{2} \cdot {\bm f}_{2} =0$) represents the motion of dust in the absence of gas. The forcing terms, ${\bm F}_{2\rightarrow 1}$ and ${\bm F}_{1\rightarrow 2}$ represent the forcing of gas motions by dust motions and the forcing of dust motions by gas motions, respectively.  RDI theory states that resonant instabilities arise from  the effect of  system one forcing system two ($\bm{\xi}^{L}_{2}\cdot \bm{F}_{1\rightarrow 2}$) multiplied by the  effect of system two forcing  system one ($\bm{\xi}^{L}_{1}\cdot \bm{F}_{2\rightarrow 1}$; see equation (4) of \citealt{Squire2018}).\footnote{In the notation 
of \citealt{Squire2018}, system 2 is $\mathcal{A}$, system 1 is $\mathcal{F}$, $F_{2\rightarrow 1}=T^{(1)}_{FA}\cdot \xi^{R}_{\mathcal{A}}$ and $F_{1\rightarrow 2}=\mathcal{C}\cdot \xi^{R}_{\mathcal{F}}$.}  If both of these 
effects are strong ($\bm{\xi}^{L}_{2}\cdot \bm{F}_{1\rightarrow 2}\neq0$ and $\bm{\xi}^{L}_{1}\cdot \bm{F}_{2\rightarrow 1}\neq0$), \revchng{the coupled system 
will exhibit exponential growth, so long as the two are not perfectly out of phase}. This makes intuitive sense: a small excitation of the first eigenmode will cause a particularly strong response in the second, which in turn causes a strong response in the first, thus leading
to runaway growth of the coupled modes. 

\begin{figure}
\begin{center}
\includegraphics[width=1.0\columnwidth]{\figfold 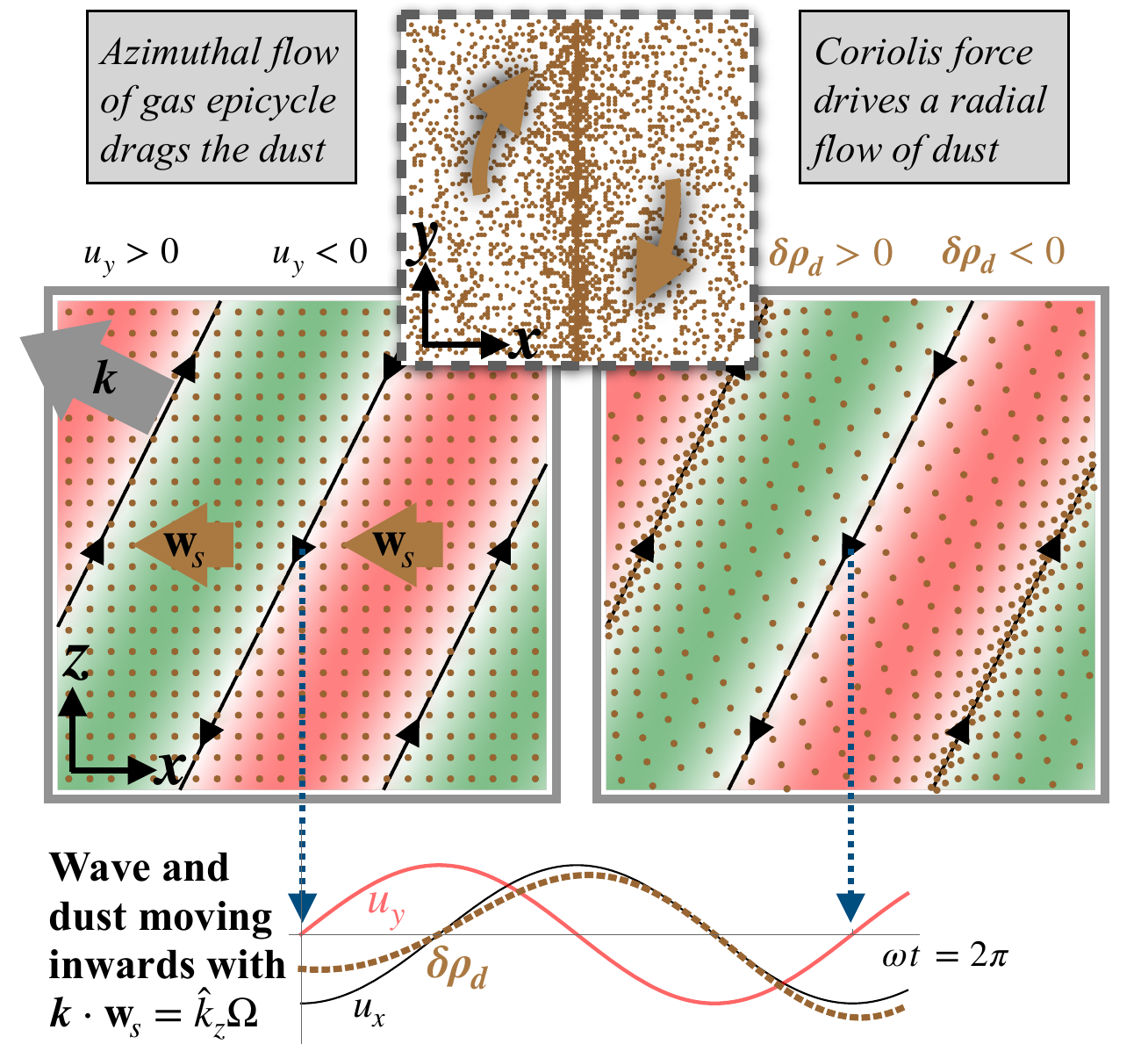}
\caption{{Response of the dust to a gas epicyclic perturbation.} The structure of a positive-frequency gas epicycle, which propagates in the $\hat{\bm{k}}$ direction, is shown in the left-hand panel. Lines show the $u_{x}$ and $u_{z}$ flow lines (crests), colours show $u_{y}$ perturbations (red  $u_{y}>0$, green $u_{y}<0$). We see that  $u_{x}$ lags $u_{y}$ by $\pi/2$ as the wave moves inwards (leftwards). The $u_{y}$ perturbations cause in-phase $v_{y}$ perturbations due to the drag force, which then generate $v_{x}$ perturbations due to the \revchng{Coriolis force} (top sub panel, \revchng{which is centred on a maximum of $u_{x}$ and $u_{z}$}). 
This compressive $v_{x}$ perturbation generates a dust-density perturbation between the regions of large $v_{y}$. If the 
natural dust frequency $\omega=\bm{k}\cdot\ws$ matches that of the epicycle, dust density 
perturbations grow secularly in time, with density maxima in phase with $u_{x}$ maxima.  \revchng{The bottom panel
illustrates qualitatively the time evolution of a point in the middle of the domain (see blue-dashed lines) over one wave period, showing the secular growth of dust-density perturbations $\delta \rho_{d}\propto t \cos(\omega t)$.}
}
%\pfh{There's a lot happening here. First lots of variables, some of which aren't defined before or only implicitly in full text. I'd add a short table before anything defining variables (${\bf u}$, ${\bf v}$, $\delta{\rho}_{d}$, etc. make it clear u and v are perturbative velocities, correct? I'd also think about including an even earlier figure, perhaps alongside the table, to just show your ``visual notation''---clearly label ``brown points = dust'', colors = gas $u_{y}$ motion, lines show phase crests/extrema of $u_{x}$ (?), etc. you try to do this here but its a LOT to process all of this in the same figure with a non-trivial mode in all of this (there's just so much presented all at once it reads a bit overwhelming} }
\label{fig:basic.gas2dust}
\end{center}
\end{figure}

\begin{figure}
\begin{center}
\includegraphics[width=1\columnwidth]{\figfold 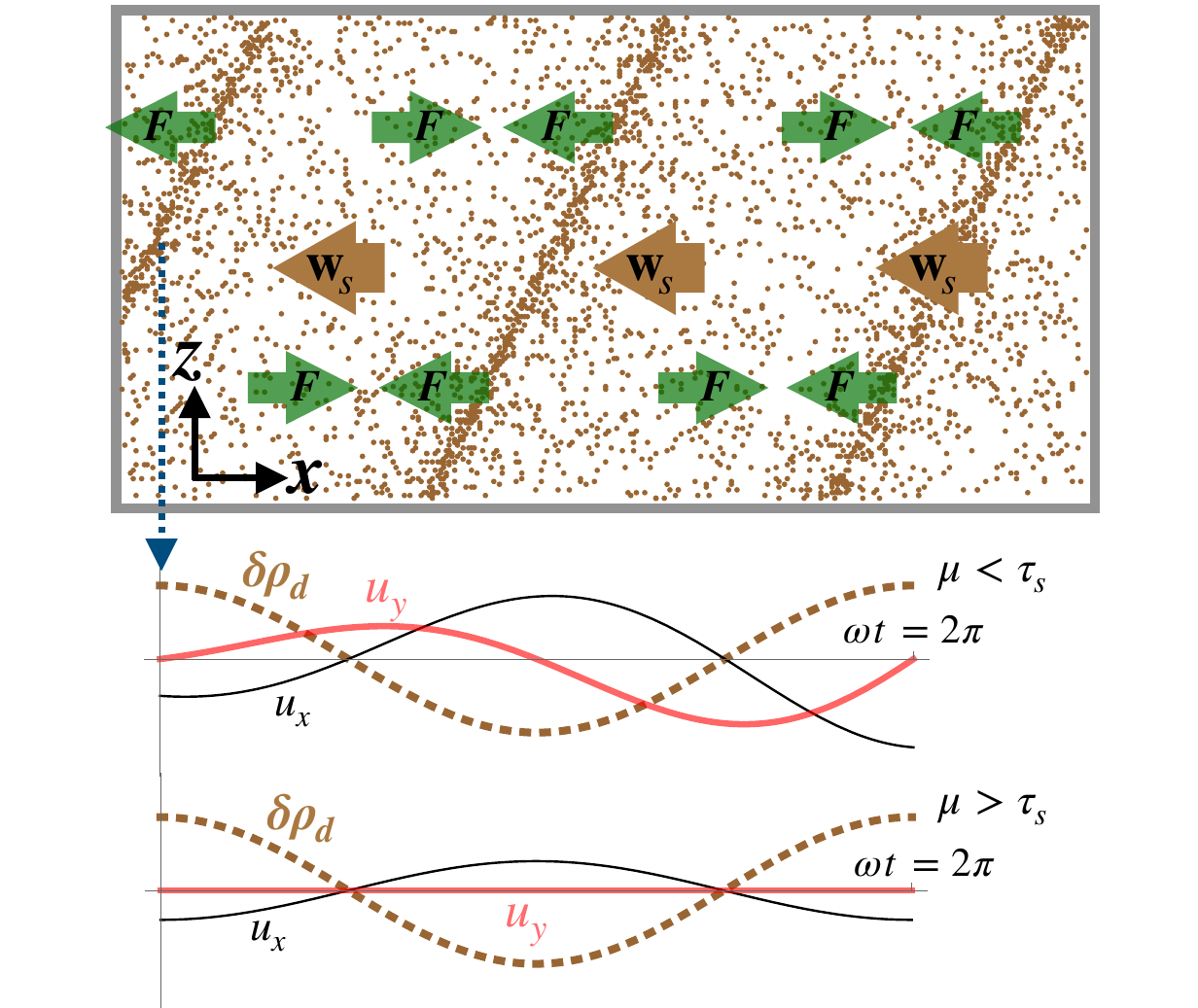}
\caption{{Response of (incompressible) gas to a dust density perturbation.} 
%The green arrows illustrate the local effective force on the gas from the dust ($\bm{F}$. 
Local dust density perturbations produce a variable force $\bm{F}$ on the gas in the direction $\delta \rho_{d}\ws$, {because the mean dust drift $\ws$ leads to the dust-density-dependent back-reaction force ${\bm F} = \delta \rho_{d} \ws/\ts$}. 
This is shown with green arrows in the top panel. This periodic
forcing with frequency $\bm{k}\cdot\ws$ excites gas epicycles if $\mu\ll \ts$, which will grow 
secularly if their frequency matches that of the forcing (top time-trace panel). If $\mu\gg\ts$, the forcing balances the drag force, directly driving a velocity in the direction $\bm{F}\propto\delta \rho_{d}\ws$ (bottom time-trace panel). When $\ws$ is parallel to $\bm{k}$, no gas motions
can be excited because the gas pressure force (incompressibility) directly opposes the dust forcing. }
\label{fig:basic.dust2gas}
\end{center}
\end{figure}

In the next two subsections and figures \ref{fig:basic.gas2dust} and \ref{fig:basic.dust2gas}, we apply this idea to 
the coupled dust-gas system in a disk. These ideas are then used in \S\ref{sec: streaming}--\S\ref{sec: high mu streaming} to explain the operation of the streaming and settling instabilities.

%%%%%%%%%%%%%%%%%%%%%%%%%%%%%%%%%%%
%\begin{figure}
%\begin{center}
%%\includegraphics[width=1.0\columnwidth]{}
%\caption{\textcolor{red}{Figure key. Show what colors and symbols refer to what. }}
%\label{fig: figure key}
%\end{center}
%\end{figure}
%%%%%%%%%%%%%%%%%%%%%%%%%%%%%%%%%%%

%%%%%%%%%%%%%%%%%%%%%%%%%%%%%%%%%%
%%%%%%%%%%%%%%%%%%%%%%%%%%%%%%%%%%

%%%%%%%%%%%%%%%%%%%%%%%%%%%%%%%%%%
%%%%%%%%%%%%%%%%%%%%%%%%%%%%%%%%%%

\subsection{The influence of gas motions on the dust}\label{sub: gas on dust}

Based on the ideas laid out in the previous paragraphs, let us consider the effect of a gas epicycle on 
the dust. This section is intended to be read in conjunction with figure~\ref{fig:basic.gas2dust}, which sketches
the relevant ideas. 
%
%\pfh{Again you're introducing some math without defining it -- the $k$ terms here are not defined. I'd just say somewhere clearly that everything in this paper will be for linearized, single Fourier modes with the usual convention $\delta X = \delta X_{0}\,\exp{(i \omega - {\bf k}\cdot {\bf x})}$. Also I don't think you ever defined the coordinate system -- you did implicitly but explicitly say somewhere cylindrical coordinates with $r$, $z$, $\phi$. }

We are interested in gas epicycles, which arise due to the Coriolis force and  local shear (see Eq.~\eqref{eq: coriolis}), and have frequency $\omega=\pm\hat{k}_{z}\Omega=\pm (k_{z}/k)\,\Omega$ for axisymmetric motions. 
The $k_{z}\neq0$ requirement for oscillations occurs because of the incompressibility condition, which implies that
the radial motions generated by azimuthal motions must also be accompanied by vertical motions. The structure of the mode, shown in figure~\ref{fig:basic.gas2dust}, involves in-phase radial and vertical motions with  azimuthal 
velocity perturbations $\pi/2$ out of phase (either leading or trailing, depending on the sign of $\omega$).

To lowest order in $\tau_{s}$, the dust simply follows the gas motions. However, in order to cause dust-density perturbations,  deviations of the dust bulk velocity from that of gas are  necessary because the gas is incompressible. The strongest of these deviations is caused by the \revchng{Coriolis force} on the dust azimuthal velocity, which generates a radial 
dust velocity $\partial_{t} v_{x} = 2 \Omega v_{y}$. In an epicycle, the $u_{y}$ perturbation generates an azimuthal dust velocity perturbation ($v_{y}$) through drag, which in turn creates a  radial dust-velocity perturbation $v_{x}$. Because this $v_{x}$ generated by the Coriolis force is not accompanied by vertical dust motions (as is the case for the $v_{x}$ caused directly by the epicycle) it is compressive, $\nabla\cdot\bm{v}\neq0$ and generates a dust-density perturbation $\delta \rho_{d}$. In the absence of dust backreaction, this forcing of the dust density perturbation 
will cause it to grow secularly in time if the dust's natural frequency matches that of the epicycle. This frequency is determined by the advection 
of the dust at velocity $\ws$ and the lack of a dust pressure response, which 
implies that the dust mode's
natural frequency is simply that of bulk motion of a static  perturbation, $\omega=\bm{k}\cdot\ws$ (in the
gas frame). Thus, an epicycle for which $\bm{k}\cdot\ws=\pm \hat{k}_{z}\Omega$ will excite strong dust 
density perturbations, which grow secularly in time, with the density maxima in phase with the gas $u_{x}>0$ perturbations. 

For a more rigorous description of the above, it is helpful to write down the velocity  components of the left eigenmode $\bm{\xi}_{D}^{L}$ associated with a dust density perturbation, which is straightforwardly derived from Eqs.~\eqref{eq: dust density}--\eqref{eq: dust NL} (see \sh) as\footnote{Note that  of $\bm{\xi}_{D,\bm{v}}^{L}$ has units of inverse velocity to give a dimensionless quantity when dotted with the right eigenmode, \emph{viz.,} it is a covector. }
\begin{align}
\bm{\xi}_{D,\bm{v}}^{L}& = \left( v_{x},  v_{y},  v_{z} \right) \nonumber \\
%&=t_{s}\left( - i \frac{k_{x}}{1+\ts^{2}}, -i \frac{2 k_{x} \ts}{1_+\ts^{2}}, -i k_{z}  \right) \nonumber \\
& = t_{s}\left(- i k_{x} ,  0, -i k_{z} \right) + t_{s} \left( i \frac{k_{x}\tau_{s}^{2}}{1+\ts^{2}}, -i \frac{2 k_{x} \ts}{1_+\ts^{2}}, 0 \right).\label{eq:dust left eig}
\end{align}
 Forcing the dust in a direction parallel to $\bm{\xi}_{D,\bm{v}}^{L}$ generates a strong dust density perturbation, while forcing perpendicular to 
$\bm{\xi}_{D,\bm{v}}^{L}$ does not cause a dust density perturbation  (see \S\ref{sub: secular.growth}). 
The first term of Eq.~\eqref{eq:dust left eig}, which is the lowest-order response in $\tau_{s}$, is simply a compressive motion, stating (as expected) that driving
a direct compression results in a strong dust density perturbation. This term is perpendicular to the
incompressible velocity field of the gas epicycle, and thus does not contribute to the dust 
density response. 
 The $y$-component of the second term relates 
to the discussion above;  a dust density perturbation can be caused by forcing a radially dependent azimuthal 
velocity. It is straightforward to verify that this term relies on the \revchng{Coriolis force} on the dust to be nonzero. The $x$-component of the second term is yet higher order in $\tau_{s}$ and will be discussed  in \S\ref{sec: streaming}.

\subsection{The influence of dust perturbations on the gas}\label{sub: dust on gas}

The excitation of gas motions due to dust-density perturbations is a simpler process. As 
discussed in point (v) of \S\ref{sub: equations} (Eq.~\eqref{eq: F.def}), because of the mean dust-gas drift $\ws$, a dust perturbation
causes an effective force per unit mass $\bm{F}=\mu\,\delta \rho_{d} \ws/t_{s}$ on the gas (compared to the equilibrium, in which  the dust force is balanced by a slight change in gas velocity; \citealp{Nakagawa1986}). This force is in the $\ws$ direction for an overdensity, or in the $-\ws$ direction 
for an underdensity. Because the
dust is advected with velocity $\ws$, this amounts to a periodic forcing 
of the gas velocity in the direction $\pm\ws$  with frequency $\omega = \bm{k}\cdot\ws$. 
This is sketched in figure~\ref{fig:basic.dust2gas}. 

The details of the gas response depend on the dust-to-gas  ratio $\mu$. If $\mu\ll \hat{k}_{z}\Omega t_{s}=\hat{k}_{z}\ts$,  gas epicycles 
are {excited}, but not modified appreciably by the dust, because the drag force is too small. In contrast, if $\mu\gg\hat{k}_{z}\ts$, 
 the Coriolis forces become {subdominant and relatively unimportant for the gas motion, because the drag dominates}. In the former 
case, the system is nearly identical to the forced oscillator  of equation~\eqref{eq:forced system}; 
the dust drives epicycles, which grow secularly in time if their frequency matches that of the dust driving. 
Just as in Eq.~\eqref{eq:forced system}, the phase of forcing ($\bm{F}\propto \delta \rho_{d}\ws$) matches 
that of the forced velocity component $(\bm{u}\cdot\ws)/|\ws|$. In the latter limit, where drag forces dominate, the gas 
velocity rapidly (faster than $\sim\!(\hat{k}_{z}\Omega)^{-1}$) reaches balance between the forcing and the drag. This again leads to 
a gas response that is in phase with the forcing from the density, although there is now no secular response 
at resonance, and the gas responds to all driving frequencies in a similar way. 
In practice, the {most} significant difference between the gas responses in {the two} regimes 
is whether or not an appreciable azimuthal gas velocity is excited by radially or vertically streaming dust: if $\mu\ll\ts$,
an azimuthal velocity $u_{y}$ is excited because gas motions are essentially epicycles; if $\mu\gg \ts$, the $u_{y}$ response is weak because the Coriolis forces are weaker than the dust-driving forces.
Nonetheless,  because the phase of the $u_{x}$ and $u_{z}$ responses remains the same in the $\mu\ll\ts$ and $\mu\gg \ts$ 
regimes, the character of the streaming and settling instabilities remains nearly independent of $\mu/\ts$.

\subsubsection{\revchng{The necessity of two-dimensional motions}}\label{subsub: needs to be 2d}
A dust-density perturbation for which $\bm{k}$ and $\ws$ are exactly parallel ($\bm{k}\times \ws=0$) will not excite any gas motions  if the gas is truly incompressible. This is because $\bm{F}$ (the dust backreaction force on the gas) is parallel to $\bm{k}$ so is purely compressive, and is therefore entirely resisted by the assumed infinitely strong pressure forces (see \S\ref{subsub: incompressibility}). As a consequence, for  radial streaming, a purely radial dust density perturbation cannot excite gas motions. In the opposite case of $\bm{k} \cdot \ws = 0$ ($k_{x}=0$ with radial streaming), the forcing is perpendicular to the motions required to drive the dust density perturbation, so cannot excite an instability. This explains why the traditional midplane streaming instability (with $\ws \propto \hat{x}$) cannot operate with a purely one-dimensional perturbation; it requires $k_{x}\neq 0$ and $k_{z}\neq0$. 
%\revchng{Note that the  commonly used analogy of the bicycle ``peloton,'' where dust in the region behind a clump feels less drag force, is relatively accurate: the local gas velocity perturbation caused by the dust backreaction discussed above locally  reduces the dust-gas velocity offset and thus the drag force. However, even the bicycle peloton requires gas velocity perturbations that are at least two dimensional, with a flow circulation to keep the air nearly incompressible. } \textcolor{red}{NEED TO CHANGE}

\revchng{With compressive gas perturbations, the above argument does not hold and fast-growing one-dimensional instabilities can indeed occur \citep{Hopkins2018}. However, in order to 
drive an instability,  a gas compression created by a dust perturbation must not ``outrun'' the dust
perturbation, since this would halt the feedback loop necessary for instability. 
Because gas compressions propagate at the sound speed, such instabilities require
dust drift velocities that approach the sound speed, which can never occur in dense regions of protoplanetary disks.
}

%%%%%%%%%%%%%%%%%%%%%%%%%%%%%%%%%%
\begin{figure*}
\begin{center}
\includegraphics[width=0.9\textwidth]{\figfold 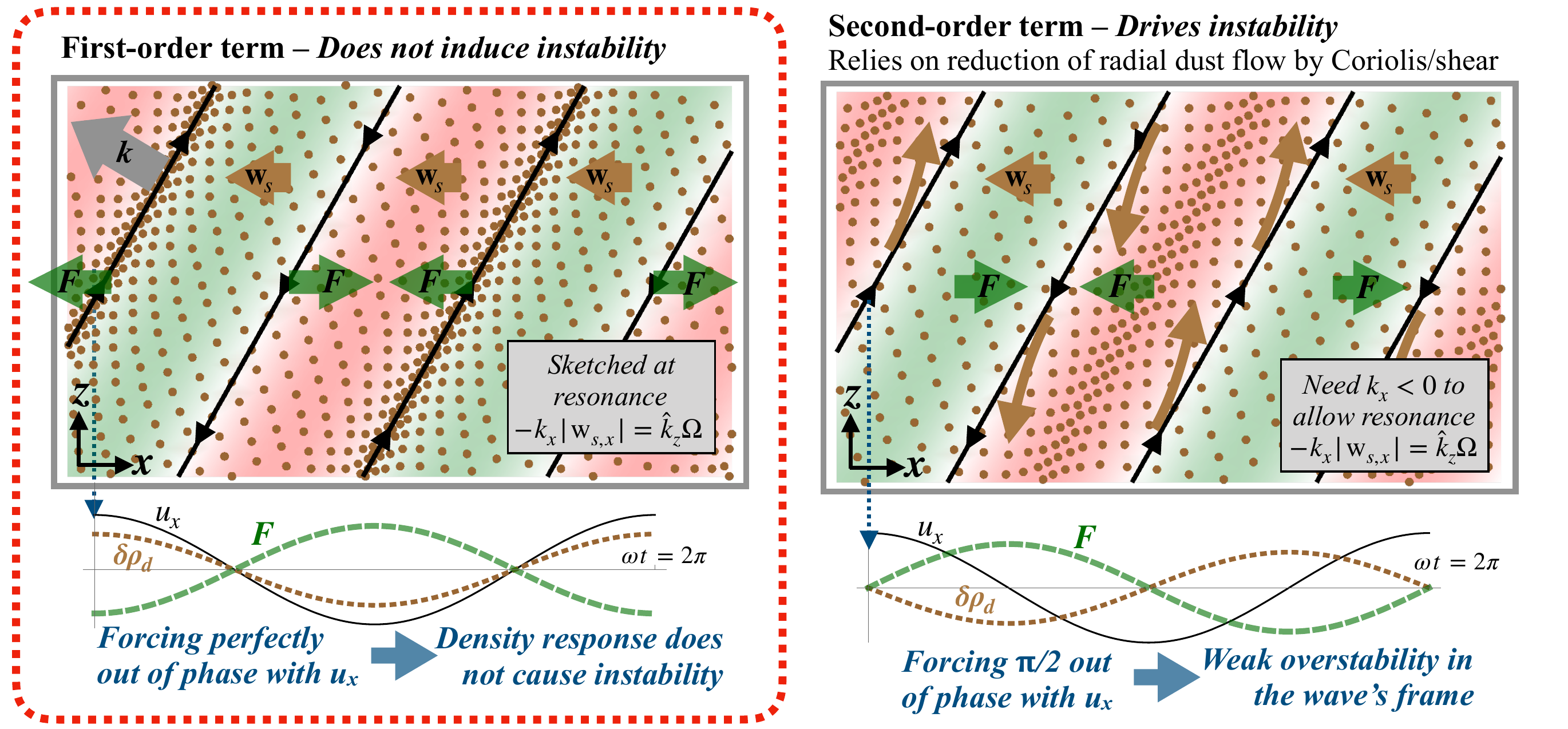}
\caption{\textbf{The low-dust-to-gas-ratio ($\mu<1$) streaming instability.} The left panel illustrates how the 
dust feedback sketched in figures~\ref{fig:basic.gas2dust} and \ref{fig:basic.dust2gas} fails to cause an instability:
the force ($\bm{F}$) from the dust density perturbation ($\delta \rho_{d}$) generated by a gas epicycle is exactly out of phase with the force that  most strongly 
amplifies the epicycle. The right panel shows the weaker interaction that does lead to instability: deflection by the
Coriliois force manifests itself as a slight decrease in the magnitude of $v_{x}$, which causes a dust compression between
maxima of $|u_{x}|$. \revchng{The velocity deflection of the dust that leads to this compression (see Eq.~\eqref{eq:dust left eig}) is illustrated with the brown arrows}. The $\pi/2$ offset between $u_{x}$ and  $\bm{F}$ generates an 
overstability in the wave frame. As in figure~\ref{fig:basic.gas2dust}, the bottom panels illustrate the time evolution 
of gas velocity ($u_{x}$), dust density ($\delta \rho_{d}$), and dust-density-induced force ($\bm{F}$) perturbations as the wave propagates in the $\bm{k}$ direction, at the point in each domain shown by a blue arrow.  }
\label{fig:streaming.lowm}
\end{center}
\end{figure*}
%%%%%%%%%%%%%%%%%%%%%%%%%%%%%%%%%%

\section{The low-$\mu$ streaming instability}\label{sec: streaming}

Through \S\ref{sec: resonances} we have seen that our understanding of instabilities should be guided
by how the dust-density perturbations generated by a gas epicycle feed back on this epicycle. In this section, we
apply these ideas to low-dust-to-gas-ratio ($\mu<1$) dust moving primarily in the negative radial direction ($\ws\propto-\hat{\bm{x}}$), as appropriate to the disk midplane. 
This is the ``low-$\mu$ streaming instability.''
Unfortunately, it transpires that its mechanism is more
complex than that of the settling instability (which involves vertically streaming dust, $\wsz\neq0$), because the dust and gas responses sketched in figures~\ref{fig:basic.gas2dust} and \ref{fig:basic.dust2gas} happen to be exactly out of phase.
 We choose to present this case first because 
it has been much better studied in previous literature, but suggest it may be helpful to a reader to consider \S\ref{sec: settling} concurrently. Helpful discussion can also be found in \citet{Zhuravlev2019} (specifically, their section 4).

The physics of the streaming instability is sketched in figure~\ref{fig:streaming.lowm}. Note that we are implicitly 
assuming that the illustrated wave satisfies the resonance condition $\bm{k}\cdot\ws = \hat{k}_{z}\Omega$, so 
that the resonant interactions sketched in figures~\ref{fig:basic.gas2dust} and \ref{fig:basic.dust2gas} are applicable.
The most important feature, shown in the left panel of figure~\ref{fig:streaming.lowm},
is that the resonant excitation of dust-density perturbations by the epicycle's azimuthal velocity 
 (figure~\ref{fig:basic.gas2dust}) has the {wrong phase to cause an instability} (the red border is used to 
 emphasise this point).
\revchng{More specifically, the interaction produces dust maxima in phase with  $u_{x}$ maxima, as shown in figure~\ref{fig:basic.gas2dust}, while 
the force from a dust maximum (minimum) is in the $-\hat{\bm{x}}$ ($+\hat{\bm{x}}$) direction (see Eq.~\eqref{eq: F.def})}. As shown in 
figure~\ref{fig:basic.dust2gas}, this force is exactly opposite to that which most strongly forces the gas, and 
thus does not cause an instability (in fact, any other phase would cause instability). In  RDI theory, 
this phase anti-alignment manifests itself mathematically as a purely real perturbation to the frequency of the epicycle at
lowest order in $\ts$ (see equation (34) of \sh).

The right-hand panel sketches the weaker dust feedback that does lead to instability. To lowest order, the 
dust follows the incompressible streamlines of the gas in the $x-z$ plane; however, due to the Coriolis forces
a small part of the dust's radial velocity is deflected into the azimuthal direction. This deflection decreases the magnitude of $v_{x}$, 
making  the flow {of dust} 
slightly compressive, and thus concentrating dust between regions of negative and positive $u_{x}$ (for a $k_{x}>0$ wave
resonating with a $\omega<0$ epicycle, the phase of $\delta \rho_{d}$ is reversed). The forcing of the gas induced by this dust density perturbation is now $\pi/2$
out of phase with the optimal driving illustrated in figure~\ref{fig:basic.dust2gas}, which  is sufficient 
to  render the epicycle unstable, causing an overstability in the wave's frame because of the $\pi/2$ phase shift.
Mathematically, this interaction is described by the $x$-component of the of the second term 
in Eq.~\eqref{eq:dust left eig} and is one order higher in $\ts$ than the compression caused by the azimuthal velocity. 
Because this term is one order higher  in $\tau_{s}$ it is outside the ``terminal velocity approximation'' for dust  (see \citealt{Zhuravlev2019} and point (iv) of \S\ref{sec: basics}).
At the same order, there is also a contribution from the azimuthal relative dust drift $\wsy$, which can modify the growth rate somewhat (\sh).

%%%%%%%%%%%%%%%%%%%%%%%%%%%%%%%%%%
\begin{figure*}
\begin{center}
\includegraphics[width=0.9\textwidth]{\figfold 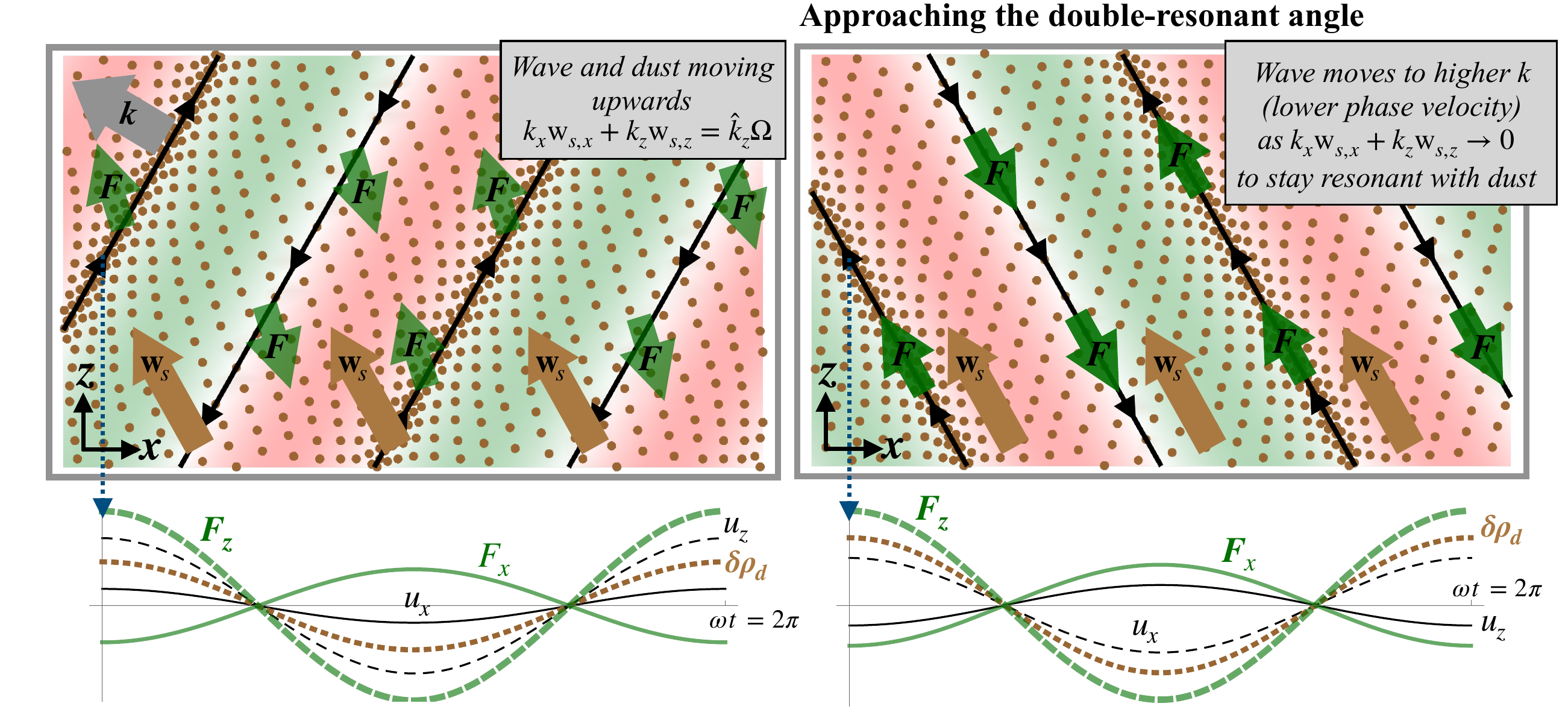}
\caption{\textbf{The settling instability.} The left panel illustrates the fundamental difference in dust-gas interaction
that arises when the dust also streams vertically: even though the radial velocity is forced out of phase by the 
dust backreaction $\bm{F}$ (as for the streaming instability; figure~\ref{fig:streaming.lowm}), the forcing of the  vertical gas velocity 
by the dust is now in phase. The resulting instability is much stronger than the streaming 
instability when $\ts<1$, because it relies on the stronger azimuthal-velocity-induced dust feedback (figure~\ref{fig:basic.gas2dust}).  The right panel 
illustrates why the ``double-resonant'' mode of \sh\ is particularly fast growing: the forcing of the epicycle by the gas is 
aligned and in phase with both $u_{x}$ and $u_{z}$, generating a particularly strong gas response that leads to growth rates 
larger than $\Omega$ at sufficiently short wavelengths (\sh). The
lower panels illustrate time traces of gas velocity ($u_{x},\,u_{z}$), dust density ($\delta \rho_{d}$), and dust backreaction force ($F_{x},\,F_{z}$) perturbations as the wave propagates in the $\bm{k}$ direction, at the point in the domain shown with the blue-dotted arrow (the dust is drifting nearly vertically with $\bm{k}\cdot\ws=\hat{k}_{z}\Omega$).  }
\label{fig:settling}
\end{center}
\end{figure*}
%%%%%%%%%%%%%%%%%%%%%%%%%%%%%%%%%%

\subsection{Model summary}\label{sub: summary streaming}

Our model  qualitatively explains  some key features of the streaming instability when $\mu<1$:
\begin{enumerate}
\item Its fastest growing modes are those that satisfy the epicycle  resonant condition $\bm{k}\cdot\ws=\hat{k}_{z}\Omega$, where the effect of gas on dust, and vice versa, is strong. This idea is explored quantitatively in \sh. 
\item The growth rate of the low-$\mu$ streaming instability decreases with $\ts$. This is not simply because 
small dust grains are ``better coupled'' to the gas, as often stated---{indeed better-coupled dust interacts
with higher-frequency motions (which might naively be associated with higher growth rates).
Rather, it is because the geometry of the system with purely radial streaming causes the normal leading-order dust backreaction force on the gas  \revchng{(Eq.~\eqref{eq: F.def})}
to be out of phase with the gas motions of an epicycle. The feedback term that causes the instability is therefore the next-order term in $\tau_{s}$, thus causing the growth rate to decrease with grain size. Equivalently, the feedback 
that drives the instability is not contained within the ``terminal velocity approximation'' \citep[see][]{Zhuravlev2019}.}
\item The streaming instability can operate only if two-dimensional motions are allowed: if $k_{z}=0$, not only does the 
gas epicycle have zero frequency (and thus cannot resonate with the streaming dust), 
but  dust perturbations cannot
feed back to excite incompressible gas motions if $\ws$ and $\hat{\bm{k}}$ are parallel. 
The general conclusion is unchanged if one allows for finite gas compressibility (see \S\ref{subsub: needs to be 2d}).
%Allowing for some gas compressibility, this condition can be relaxed, but $k_{z}\rightarrow 0$ modes will still be suppressed by finite gas pressure.
\item Dust perturbations are approximately $\pi/2$ out of phase with those of $u_{x}$ and $u_{z}$, and thus 
in phase with $\pm u_{y}$ (with the sign depending on the sign of $k_{x}$).
\end{enumerate}

\section{The settling instability}\label{sec: settling}

We now apply the same analysis as above to the case where dust is settling towards the midplane, 
with a drift velocity $|\wsz|\gg |\wsx|\gg |\wsy|$. This leads to  the ``settling instability'' of \sh. As mentioned 
above, the situation is simpler in this case than for the streaming instability because 
there is no longer a phase anti-correlation between the dust feedback and the epicycle. Properties of the
settling instability are explored extensively in \citet{Zhuravlev2019}.

As in \S\ref{sec: streaming}, our first assumption is that the resonance condition 
is satisfied, 
\begin{equation}
k_{x}\wsx + k_{z}\wsz = \pm \hat{k}_{z}\Omega,\label{eq: settling resonance condition}
\end{equation}
such that there are strong interactions between the gas and dust modes. In order to use 
the same $\omega>0,\,k_{x}<0,\,k_{z}>0$ convention as in figures~\ref{fig:basic.gas2dust}--\ref{fig:streaming.lowm} 
for illustration, let
us take $\wsz>0$, as applicable below the disk midplane. We sketch this case in the left panel of figure~\ref{fig:settling}. The dust density perturbation generated 
by the epicycle is again in phase with $u_{x}$ as shown in figure~\ref{fig:basic.gas2dust}; however, 
the force on the gas, which is proportional to $\delta \rho_{d}\wsh$, is now primarily in the $+\hat{\bm{z}}\delta \rho_{d}$
 direction. Although the $x$-directed force is again out of phase with that needed to drive the epicycle, 
 the stronger $z$-directed force is in phase with $u_{z}$, and thus will 
 strongly drive the epicycle (see figure~\ref{fig:basic.dust2gas}).
 
Modes with the opposite $k_{x}$ (for the same sign of $\wsz$) feel an even stronger feedback 
from the dust, because the dust forcing on the epicycle can be in phase with both $u_{x}$ and $u_{z}$ (see lower-right-hand panel of figure~\ref{fig:settling}). Such 
modes have somewhat smaller wavelengths because the $k_{x}\wsx$ and $k_{z}\wsz$ have opposite 
signs, thus increasing the wavenumber $k$ ($k=|\bm{k}|$) required to satisfy Eq.~\eqref{eq: settling resonance condition}. 
A particularly extreme version---in which $\ws$ is nearly aligned with the wavefronts---is sketched in the right panel 
of figure~\ref{fig:settling}. This was termed the ``double-resonant mode'' in \sh. We see that as the epicycle propagates 
(to the right and up now, because $k_{x}>0$), the force from the dust perturbation that it generates is very close to the direction 
of the incompressible gas velocity, causing a particularly strong response and  a fast growing mode. In fact, as shown in 
\sh, the growth rate of this mode approaches infinity as $k\rightarrow \infty$ for any $\tau_{s}$ and any $\mu$ ($\omega\propto k^{1/3}\mu^{1/3}\ts^{1/3}$), surpassing $\Omega$ for 
sufficiently small wavelengths \revchng{(note that  $\hat{\bm{k}}\cdot\ws\rightarrow 0$ for this mode, so the resonance condition \eqref{eq: settling resonance condition} implies that the mode must be at high $k$)}. 
As it moves to shorter wavelengths, the wave loses its epicyclic character, although the basic phase structure 
remains similar, with $\delta \rho_{d}$ being primarily driven by azimuthal velocities and the \revchng{Coriolis force}.

\subsection{Model summary}\label{sub: summary settling}

Our model  qualitatively explains  some of the key features of the settling instability:
\begin{enumerate}
\item Unlike the streaming instability, its maximum growth rate does not depend on $\tau_{s}$, and is just as large for
arbitrarily small grains ({although the characteristic resonant wavelength  decreases with smaller $\tau_{s}$}). The reason is that the dust feedback, which forces the fluid in the direction $\delta \rho_{d}\wsh$, is partially in phase with the  velocity of the epicycle. This implies that the leading-order dust backreaction force, which enables the instability, is independent of $\tau_{s}$, causing a $\tau_{s}$-independent growth rate.
\item For the same $\tau_{s}$, the settling instability operates at larger wavelengths than the streaming instability, because the vertical
settling drift is larger than the radial drift in  a thin disk (\sh).
\item Settling instability resonant modes for which the sign of $k_{x}\wsx$ is opposite to that of $k_{z}\wsz$ grow faster than those with the opposite
polarity, because the dust forcing is in phase with both $u_{x}$ and $u_{z}$ perturbations. (These modes have smaller wavelengths, however, compared to resonant modes for which $k_{x}$ and $k_{z}$ have the same sign.)
\item The ``double-resonant mode,'' which has very large ($>\!\Omega$) growth rates at small  scales, occurs 
because the forcing of the mode by $\delta \rho_{d}$ perturbations (in the direction $\bm{F}\propto\delta \rho_{d}\ws$) aligns  with the incompressible flow velocity, thus 
driving the gas motions particularly efficiently.
\item The settling instability is strongly suppressed (similar to the streaming instability) if $\bm{k}$ and $\ws$ are parallel, or, more generally, if $k_{z}/k_{x}\leq\wsz/\wsx$. In this case, the forcing is perpendicular to that required for incompressible motions (or out of phase), so only the weaker higher-order dust feedback  
of the streaming instability is effective at amplifying the epicycle (see figure~\ref{fig:streaming.lowm}, right-hand panel).
\end{enumerate}

%%%%%%%%%%%%%%%%%%%%%%%%%%%%%%%%%%
\begin{figure}
\begin{center}
\includegraphics[width=1.0\columnwidth]{\figfold 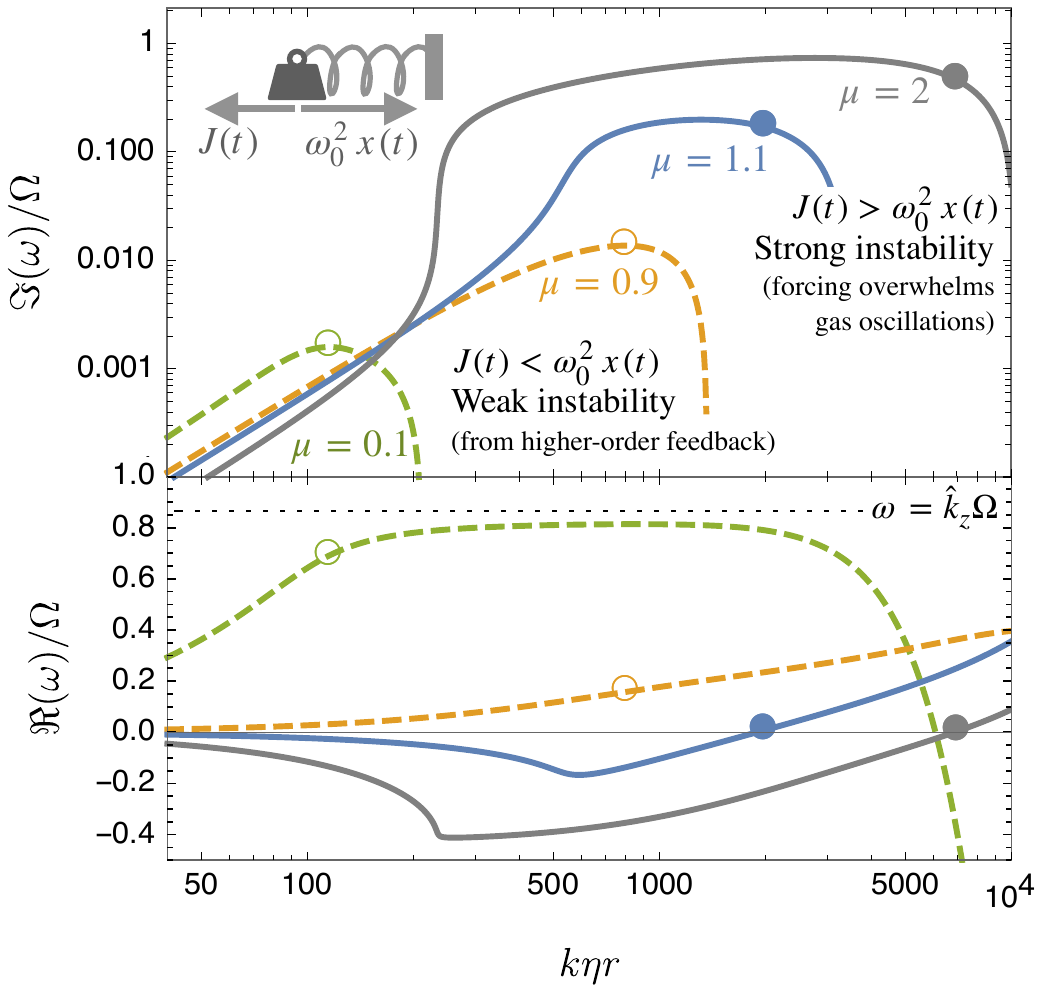}
\caption{{Illustration of the  sudden change to the streaming instability  at ${\mu=1}.$} We show the growth rate (upper panel) and
oscillation frequency (lower panel) of modes as a function of wavelength, for $\tau_{s}=0.01$ and $\mu=0.1$, $\mu=0.9$, $\mu=1.1$, and $\mu=2$. Each curve is calculated from the full dispersion relation as a function of $k\eta r$ (see \sh), with a mode angle of $30^{\circ}$ ($k_{z}/k_{x}=\sqrt{3}$; other mode angles have similar behavior).
At $\mu<1$, shown with dashed curves, the resonant (fastest-growing) wavelength of the streaming instability is 
effectively an epicycle, modified by the force $\bm{F}$ on the gas from the dust.
Its real oscillation frequency  decreases steadily from its unperturbed value ($\omega=(k_{z}/k)\Omega$) as $\mu$ increases (unfilled circles label the resonant $k$ and $\omega$ in both panels). At $\mu>1$, shown with solid curves, the dust forcing overwhelms the oscillator ($\bm{F}$ is stronger than the restoring epicyclic force) causing the mode to become nearly purely growing and changing the phase relationship between components of $\bm{u}$. This produces a sudden increase in the maximum growth rate. The zero-frequency mode, where the forced-oscillator analogy in the text (\S\ref{sec:out.of.phase.forcing}) is most appropriate, is labeled with filled circles in each panel. The transition---i.e., the change from a mode with $\Re(\omega)>\Im(\omega)$ to one with $\Re(\omega)<\Im(\omega)$  at $\mu=1$---is well understood using the forced-oscillator analogy, while the behaviour of the fast-growing modes with $\mu > 1$ is explained by the model in figure~\ref{fig:streaming.highm}.}
\label{fig:gamma}
\end{center}
\end{figure}
%%%%%%%%%%%%%%%%%%%%%%%%%%%%%%%%%%

%%%%%%%%%%%%%%%%%%%%%%%%%%%%%%%%%%
\begin{figure*}
\begin{center}
\includegraphics[width=1\textwidth]{\figfold 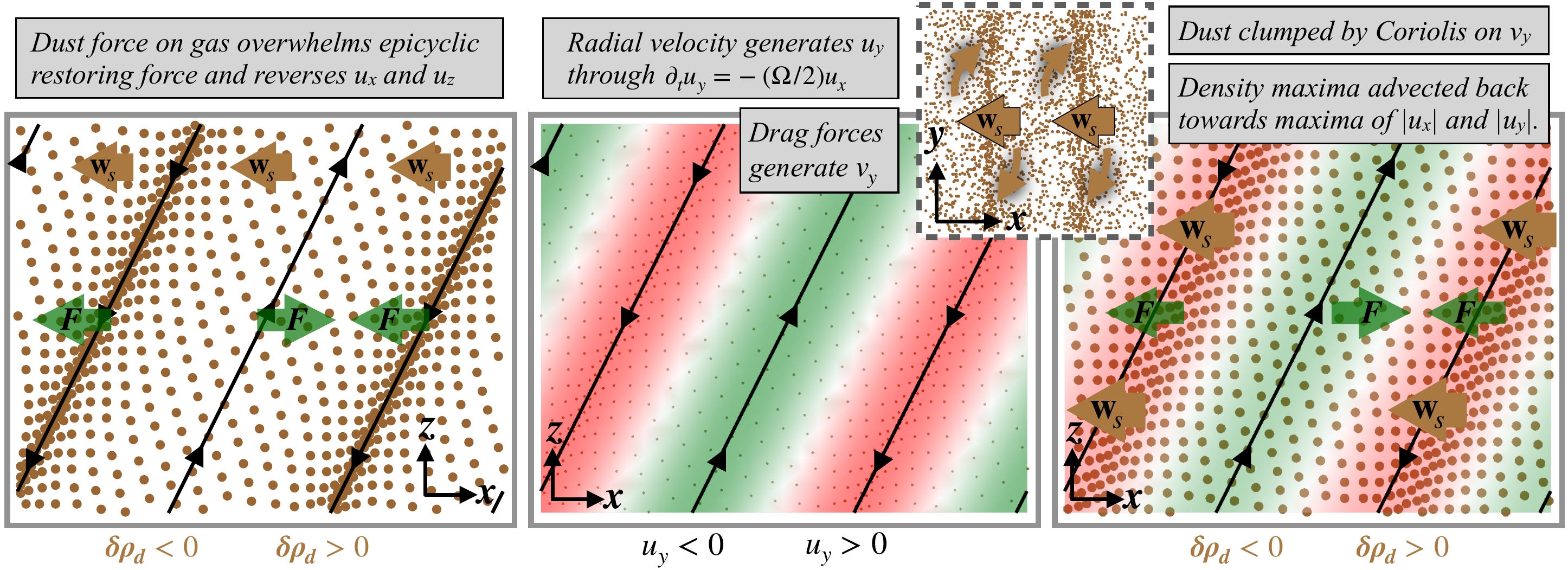}
\caption{\textbf{The high-dust-to-gas-ratio ($\mu>1$) streaming instability}. For $\mu>1$, the force of the dust density perturbation on the gas ($\bm{F}$)
overwhelms the natural restoring force of the epicycle (left panel). This reverses the sign of the radial gas velocity (compare to figure \ref{fig:basic.gas2dust}), which in turn 
generates an azimuthal velocity through the Coriolis/shear force (middle panel). Note the different phase relationship  
between $u_{x}$ and $\delta \rho_{d}$, and between $u_{y}$ and $u_{x}$ compared to the unmodified epicycle shown in figure~\ref{fig:basic.gas2dust} (red and green regions show $u_{y}>0$ and $u_{y}<0$, respectively). This azimuthal velocity causes a radial compression
of the dust density (see top inset and figure~\ref{fig:basic.gas2dust}), which is advected inwards by the dust's drift to be back in
phase with the original perturbation, thus causing instability. Note that, unlike the sketches in figure~\ref{fig:streaming.lowm} and the left panel of figure~\ref{fig:settling}, the pictured mode is stationary (purely growing) or slowly propagating outwards (which allows the density perturbation to stay in phase with the velocity perturbations at longer wavelengths).  }
\label{fig:streaming.highm}
\end{center}
\end{figure*}
%%%%%%%%%%%%%%%%%%%%%%%%%%%%%%%%%%

\section{The high-$\mu$ streaming instability}\label{sec: high mu streaming}

A key feature of the streaming instability---noted in \citet{Youdin2005} and \citet{YoudinA2007} and studied in more detail in
Appendix~A of \sh---is a sudden change in its growth rate and other properties as the dust-to-gas ratio $\mu$ increases 
beyond unity. The transition, which is  illustrated in figure~\ref{fig:gamma} and occurs for any $\tau_{s}\lesssim1$, is likely a key feature of the streaming instability's \emph{nonlinear}
development, since simulations seem to show that the presence of  high-$\mu$ regions in the midplane 
is a necessary requirement for the formation of gravitationally-bound planetesimals.
We discuss this ``high-$\mu$ streaming instability'' in this section, illustrating how it operates, why it grows so 
much more rapidly than
the $\mu<1$ streaming instability, and why it  
appears only for $\mu\geq1$.
As a reminder, our discussion here will not be mathematically rigorous, rather focusing on providing intuitive insight. 
We will, however, occasionally use expressions from Appendix~A of \sh\ for clarity. Similarly, figure~\ref{fig:gamma} illustrates 
streaming instability growth rates calculated from a full (numerical) solution to the dispersion relation.
%, illustrating thesudden change to $\Im(\omega)$ at $\mu=1$.

\revchng{Note that, unlike the streaming instability, there is not a sudden change in the properties of the settling instability 
as $\mu$ surpasses unity, although its growth rate does keep increasing with $\mu$ with fast-growing modes that resemble 
the double-resonant mode. For this reason, and because it seems less astrophysically relevant---dust is unlikely to reach such high densities outside of the midplane in disks---we do not discuss the settling instability at $\mu>1$.}

\subsection{Out-of-phase forcing of the epicycle}
\label{sec:out.of.phase.forcing}

In \S\ref{sec: streaming} and figure~\ref{fig:streaming.lowm}, we saw
that if the dust is streaming radially, the  dust-density perturbation induced by an epicycle  is almost perfectly 
out of phase with the perturbation needed to create a strong response from the gas.
Only a much weaker (higher order in $\ts$) dust response feeds back to the gas and
causes instability, explaining the decrease in streaming instability growth rate with $\ts$.
Here we discuss how the  lowest-order  feedback does nonetheless have important effects: it is responsible 
for the streaming instability's character change at $\mu=1$.

A useful rough analogy for the lowest-order out-of-phase feedback 
is that of a spring with a external forcing that is proportional to, but in the opposite direction to, its displacement:
\begin{equation}
\frac{\partial^{2}}{\partial t^{2}}x(t) + \omega_{0}^{2} x(t) = J(t),\quad J(t) = J_{0} x(t).\label{eq: toy oscillator high mu}
\end{equation}
Here $x(t)$ is the spring's displacement, which represents the radial velocity of the epicycle, and $J(t)$ is
the external forcing, which represents the jerk (rate of change of acceleration) from the dust feedback. 
The frequency of the oscillator, $\omega = \pm\sqrt{\omega_{0}^{2} - J_{0}}$, decreases as 
the forcing becomes comparable to the restoring force,  splitting into purely growing and decaying modes  once $J_{0}>\omega_{0}^{2}$. This transition accounts for the step 
change in streaming instability properties at $\mu=1$: when the dust mass density is equal to that of the gas,
 its forcing on the gas becomes larger than the restoring force of the epicycle,
 flipping the sign of the radial gas velocity and causing a near purely growing mode. Indeed,
as shown in \sh, the frequency of the fastest-growing streaming instability modes for $\mu>1$ are  simply $\omega \approx \pm i\Omega\sqrt{\mu-1}$,
exactly as expected from this forced-oscillator analogy.\footnote{Note that the fastest-growing
modes occur when $k_{z}\gg k_{x}$, so that the epicyclic frequency---$\omega_{0}$ in Eq.~\eqref{eq: toy oscillator high mu}---is simply $\Omega$ (i.e., $\hat{k}_{z}\approx 1$). As shown in \sh, in the
opposite limit $k_{x}\gg k_{z}$ the growth rate is reduced to $\hat{k}_{z}\Omega\sqrt{\mu-1}$, also exactly 
as expected from our simple oscillator analogy.} 
In addition, the  analogy predicts that the real frequency of the resonant 
mode at $\mu<1$ should steadily decrease with increasing $\mu$. Figure~\ref{fig:gamma}---which shows 
the growth rate and real frequency of the relevant modes at $\tau_{s}=0.01$ across the $\mu=1$ transition---confirms that 
this is the case, as well as showing the sudden increase in $\Im(\omega)$ at $\mu=1$ (recall that at $\mu<1$, the mode is weakly growing only because of the 
higher-order dust interaction shown in figure~\ref{fig:streaming.lowm}, which is not included in the forced-oscillator analogy).

\subsection{Wavenumber dependence}

Our simple analogy neglects the mode's $k$ dependence, assuming that the 
compression of the dust by the epicycle and the force of the  dust on the gas remain synced (in resonance). 
To understand the wavenumber dependence, we must  consider, in addition to the epicycles, 
 how the natural frequency of 
the dust density advection changes as $\mu$ increases.\footnote{Mathematically, this feedback can be worked
out using perturbation theory on the dust-density eigenmode.}
Because of the feedback force on the gas  $\bm{F}=\delta \rho_{d}\ws/t_{s}$, 
an advected dust-density perturbation drives radial gas flows as it propagates. These, in turn, drive a compressive 
dust flow due to the \revchng{Coriolis force}, which is out of phase with the advection term $\ws\cdot\nabla\rho_{d}$. 
The net effect is to slow
the advection of the dust density perturbation by a factor  $(1-\mu)$. 
Comparing the dust-modified epicycle frequency $\omega\approx(1-\mu)^{1/2}\hat{k}_{z}\Omega$ 
\revchng{(from Eq.~\eqref{eq: toy oscillator high mu})} to the 
dust-advection frequency, $\omega \approx (1-\mu) \bm{k}\cdot\ws$, 
we see that the two are matched at a  wavenumber that scales as  $k\propto(1-\mu)^{-1/2}$, formally approaching infinty 
as $\mu\rightarrow 1$. This scaling of the fastest-growing wavenumber is indeed seen 
in numerical solutions of the full dispersion relation of the $\mu<1$ streaming instability (see $\mu<1$ solutions in figure \ref{fig:gamma}). 
More importantly, we see that the $\mu>1$ instability occurs at very short wavelengths, where physics
beyond our simple toy analogies comes into play. Indeed, as shown in \sh, the instability's fastest-growing 
modes increase faster with decreasing $\tau_{s}$ than those of the low-$\mu$ instability.\footnote{Specifically, equations~(A2)--(A5) of \sh\ shows that the fastest growing wavenumber $k_{\rm max}$ behaves as $k_{\rm max}\!\sim \tau_{s}^{-5/4}$ to $k_{\rm max}\!\sim \tau_{s}^{-3/2}$, depending on the regime, as opposed to $k_{\rm max}\!\sim \tau_{s}^{-1}$ for the low-$\mu$ streaming or settling instabilities.}

\subsection{Mode structure and growth}
A sketch of how the high-$\mu$ streaming instability operates is shown in figure~\ref{fig:streaming.highm}. 
An important distinction compared to our sketches of the low-$\mu$ streaming and settling instabilities (figures~\ref{fig:streaming.lowm} and \ref{fig:settling}) is that the mode in the sketch is stationary (purely growing), or slowly propagating to the right (opposite to figure~\ref{fig:streaming.lowm}).
With $\mu>1$, the force from the dust perturbation generated by the azimuthal velocity (see figure~\ref{fig:basic.gas2dust})
is sufficiently large to flip the sign of the radial gas velocity, aligning dust-density maxima with \emph{negative} radial velocities (this 
is reversed from figure~\ref{fig:basic.gas2dust}).
This radial velocity, in turn, generates an azimuthal gas and dust velocity through the Coriolis/shear force. Because 
$\partial_{t}u_{y} = -\Omega/2\,u_{x}+\dots$ and 
the mode is dominated by its exponential growth rather than oscillations, $u_{y}$ has negative maxima near the 
positive $u_{x}$ maxima, and vice versa.  Note that the \revchng{Coriolis force} on the radial gas 
velocity ($\partial_{t}u_{x}=2\Omega u_{y}+\dots$), which would tend to produce radial velocities of the opposite sign,
is dominated by the dust feedback (as must be the case for the forced-oscillator analogy to be correct).

The \revchng{Coriolis force} on the radial \emph{dust} velocity, by contrast, remains important, because this generates the
necessary compressive flow to cause the dust-density perturbation that  sustains the mode (top inset; see figure \ref{fig:basic.gas2dust}). 
In the absence of dust advection, or if the mode were propagating in sync with the density advection (as in figure~\ref{fig:streaming.lowm}), this density perturbation would be $\pi/2$ out of phase with $u_{y}$ ($\delta \rho_{d}$ 
maxima to the right of $u_{y}$ maxima; see top inset); 
however, the background dust drift causes perturbations to be advected inwards at the same 
time as they are produced,
bringing them back into alignment with the negative radial velocity and feeding back coherently on the
original mode. As can be seen in figure~\ref{fig:gamma} (blue and grey curves), at longer wavelengths, 
the mode propagates slowly  outwards ($\Re(\omega) <0$), against the drift, which allows the dust density perturbation to stay 
in phase with $u_{x}$ by reducing the necessary inwards shift in $\delta \rho_{d}$ due to advection. At wavelengths that are
shorter than that where the mode is purely growing ($\Re(\omega)>0$) the mode is quickly killed because the density perturbation
would be advected too far inwards to cause the necessary feedback on the gas.  Similarly, as $\mu$
increases, the increase in mode growth rate  due to the increased dust force on the gas ($\omega\propto\!\sqrt{\mu-1}$)
requires that the mode move to shorter wavelengths 
because the dust density perturbation must be 
advected back across the mode wavelength more rapidly  (in addition, $\ws$ decreases with $\mu$, which exacerbates the effect).
This move to shorter wavelengths at increasing $\mu$ can be  seen in the dispersion relation shown in figure~\ref{fig:gamma} (compare $\mu=1.1$ and $\mu=2$ solutions).
%Detailed mathematical scalings are derived in \sh.

%\pfh{This is nice, but the last statement is a bit abrupt (makes the reader ask ``why not simply read paper 1?''---Not sure quite how to rephrase---maybe say more exact mathematical expressions/expansions justifying the above are in paper one, but not ``put together'' or consequences like what you have below not presented there?}
% Have discussed this connection lots earlier now, so it didn't seem necessary anyway

\subsection{Model summary}\label{sub: summary streaming highm}

Our model qualitatively explains the following important features of the streaming instability:
\begin{enumerate}
\item The sudden increase in the growth rate of the streaming instability at $\mu=1$, which is when the 
forcing on the gas from the dust  overwhelms the natural restoring force of epicyclic oscillations.
\item The maximum growth rate of the streaming instability is independent of grain size $\tau_{s}$ if $\mu>1$  (for $\ts\lesssim1$).
\item The presence of a purely growing streaming instability if $\mu>1$. (Or, at somewhat longer wavelengths, 
the streaming instability's reversed propagation direction compared to the $\mu<1$  instability.) The highly simplified model can even  
predict the mode's fastest growth rate, $\omega \approx i \Omega \sqrt{\mu-1}$.
\item The short wavelength of the high-$\mu$ instability, and the decrease of this wavelength with $\mu$. The mechanism requires that a dust 
density perturbation be advected across the mode sufficiently rapidly so that the
force it induces on the gas aligns with the radial gas velocity. At larger $\mu$, the faster growth rate  
requires that this happens more rapidly, necessitating shorter wavelengths.
\item The high-$\mu$ streaming instability grows fastest when  $k_{x}\ll k_{z}$, because the dust forcing on the gas (in the $\hat{\bm{x}}$ direction) is closely aligned with the incompressible velocity (in direction $\pm \hat{k}_{z}\hat{\bm{x}}\mp  \hat{k}_{x}\hat{\bm{z}}$). However,  $k_{x}\ll k_{z}$ modes also necessitate smaller wavelengths, because 
short radial wavelengths are required for operation of the mode (see  point (iv)).
\item Both inwards dust drift and \revchng{Coriolis forces}  are necessary for the mechanism of the high-$\mu$ streaming instability. 
Dust drift 
exerts a force on the radial gas velocity and shifts  density perturbations back in phase with the mode. 
 Coriolis/shear forces generate azimuthal from radial velocities, and in turn, dust density 
 perturbations from the  azimuthal dust velocity perturbations.
\item The gas and dust velocities in the high-$\mu$ streaming instability do not have the same structure as epicyclic oscillations. 
Indeed, the mode is very different in character to an epicycle, unlike the $\mu<1$ streaming instability.
\end{enumerate}
\revchng{Finally, the model also explains how the  wavelength of the  fastest-growing low-$\mu$ ($\mu<1$) streaming instability mode decreases as $\mu$ increases.
This occurs because the dust advection frequency decreases more rapidly with $\mu$ than the epicyclic frequency, meaning the effective resonant wavelength decreases.  }

\section{Conclusions}

This article is intended to  elucidate the key behaviours and origins of three related dust-drag induced instabilities: (i) the  dust-gas
 ``streaming instability''  at low dust-to-gas mass ratios ($\mu < 1$; \citealp{Youdin2005}); (ii) the disk ``settling instability'' of \sh\  \citep{Squire2018a}; and (iii) the  streaming instability at
 high dust-to-gas-ratios ($\mu>1$). By facilitating the coagulation of grains from the smallest dust into larger 
gravitationally bound objects, these instabilities are believed to play a key role  in planetesimal formation in protostellar disks. 
Each instability  derives from the combination of dust drift, rotation 
(\revchng{Coriolis} forces), and gas pressure, while the low-$\mu$ streaming and settling instabilities are  in the family of 
``resonant drag instabilities'' \citep[RDIs;][]{Squire2018}. However, each of the three  also exhibit fundamental and qualitative differences in their  physical driving mechanisms, which strongly influences  their mode structure and growth rates.

%We combine exact structure of modes and analytic results derived in \sh\ with simple analytic toy models,'' and detailed illustrations of the behavior of gas and dust modes in various limits of the instabilities, in order to better understand and illustrate the actual dynamics of the instabilities and how they operate differently in different regimes (and the basic physics that drives their key behaviors). Our philosophy has been to eschew detailed mathematical derivations (all of which are presented in \sh) in favour of a combination of the simplest-possible representations of the dynamical equations (together with detailed figures) that provide a mostly-complete picture of the operation of the instabilities. 

Throughout the article, our philosophy has been to eschew detailed mathematical derivations in favour of representing 
the crucial features of the dynamical equations in the simplest way possible. Along with detailed figures 
that sketch the key motions and forces involved in each mode, these simplified models enable 
a straightforward, intuitive understanding of the instabilities' operation that has been lacking in previous literature.  
Further, combined with the detailed analytic derivations of growth
rates and mode structure presented in \sh, they provide a relatively complete 
picture of the streaming and settling instabilities' operation and how their properties change with
parameters.

%All of these are resonant drag instabilities of the general family defined in \citet{Squire2018}, and all derive from a combination of dust drift in a rotating system with non-zero Coriolis/shear force and non-negligible pressure. Yet despite this universal structure, there are fundamental and qualitative differences in the behaviors and detailed physical driving mechanisms in different regimes (e.g.\ low-vs-high $\mu$, or small-vs-large grains). 

Our models are summarised in figures~\ref{fig:streaming.lowm}, \ref{fig:settling}, 
and \ref{fig:streaming.highm}, which apply, respectively, to the low-$\mu$ streaming instability, the settling instability, and the high-$\mu$ streaming instability. These figures are designed to be digestible without detailed reference to the text (see table~\ref{tab:} for definitions and conventions). In addition, 
\S\ref{sub: summary streaming}, \S\ref{sub: summary settling}, and \S\ref{sub: summary streaming highm} provide a summary of the key features of each regime that are explained by our models. Some
of the most important of these conclusions are:
\begin{enumerate}
\item \textbf{Dust clumping due to the Coriolis force: }Gas motions clump the dust because of the action of the \revchng{Coriolis force}. Azimuthal dust velocities are 
deflected into the radial direction by the Coriolis force, which generates a compressive radial dust flow even if the
gas motions are incompressible (see figure~\ref{fig:basic.gas2dust}).
\item \textbf{Dust backreaction from density perturbations: }Dust density perturbations feed back on the gas due to the dust's mean  drift. Regions of higher dust density   
exert a higher-than-mean force on the gas in the drift direction due to drag, while regions of lower dust density exert a 
lower force (see figure~\ref{fig:basic.dust2gas}). This feedback allows  dust clumps to re-enforce the gas motions that caused them. 
\item \textbf{Resonant modes: }When gas motions dominate, at $\mu<1$, the instabilities grow fastest at the ``resonant'' wavelength, where the
drift speed of the dust matches the phase velocity of the wave (as formalised by RDI theory; \citealp{Squire2018}). At this wavelength,
a propagating gas epicycle causes strong secular growth of dust-density perturbations, because the driving of the dust matches its
natural advection rate.
\item \textbf{Strong feedback from small grains: }\revchng{ Instabilities caused by small grains ($\tau_{s}\ll1$) generically grow just as rapidly as those caused by large grains ($\tau_{s}$ approaching 1). Smaller grains, although they are better coupled to the gas, also 
exert larger forces on the gas for the same drift speed. They thus interact more strongly with small-scale gas motions 
and cause fast-growing instabilities with short wavelengths. This conclusion applies to grains that are arbitrarily small, until a nonideal
effect such as viscosity starts to impact mode structure.
However, the  streaming instability with $\mu<1$, which is well known to have a growth rate that decreases linearly with $\tau_{s}$ at $\tau_{s}<1$ \citep{Youdin2005},  is unusually weak for small grains.  }
Its low growth rate occurs because, when dust is drifting in the disk midplane (radially),
the natural force feedback of a dust-density perturbation on a gas epicycle (point (i)) is nearly out of phase with the gas velocity of the epicycle. This neutralises  the leading-order backreaction term, which would otherwise drive an instability. This neutralisation does not occur if the dust has  non-vanishing vertical drift (the settling instability), accounting 
for the settling instability's larger growth rate, which is independent of grain size for smaller grains. 
\item \textbf{A different streaming instability at $\mu>1$: }\revchng{The streaming instability undergoes a sudden change in its properties at $\mu=1$, exhibiting
a much larger maximal growth rate that does not decrease with grain size (see figure~\ref{fig:gamma}).}
This arises because the force feedback of the dust on an epicycle, which is naturally out of phase with the 
epicycle's velocity (point (iv)), becomes so strong that it overwhelms the
natural restoring force, flipping the sign of the radial gas velocity. 
In its simplest form, the instability 
 does not propagate and does not resemble the mode structure of an epicycle because it is 
dominated by forces from the drifting dust. The instability grows rapidly compared to the low-$\mu$
streaming instability, with a maximum growth rate $\Im(\omega)\approx\Omega \sqrt{\mu-1}$, although it operates only at very
small scales for small grains. 
\item \revchng{\textbf{Pressure forces: }Gas pressure forces always dominate over dust backreaction forces (\S\ref{subsub: incompressibility}), implying gas motions are nearly incompressible. This further implies that a dust density perturbation that 
varies only in the direction of the drift ($\bm{k}$ parallel to $\ws$) cannot drive instability, because 
the backreaction force from the dust is resisted by gas pressure forces (see \S\ref{subsub: needs to be 2d}). }
\end{enumerate}
Although the basic ideas of conclusions (iii)--(vi) have appeared in various forms in  previous works (\sh; \citealp{Youdin2005,YoudinA2007,Jacquet2011,Zhuravlev2019,Jaupart2020}),
the novel feature of this work is the explanation of \emph{why} these properties arise from 
the basic physics of epicyclic motion, gas pressure forces, and dust-gas drag forces. 
Although it transpires that some of this physics is less straightforward than for other astrophysical fluid instabilities, 
 the basic ideas could be useful in future work for applications including the analysis/interpretation of nonlinear simulations  and 
 the development of nonlinear models of streaming-instability-induced turbulence. 

%A number of possibilities exist for building on the models in this work. The recent work of \citet{Krapp2019}
%illustrates the  importance  of including a range of grain sizes  in studying the streaming 
%instability, particularly at $\mu<1$. While our focus here has been on distilling out the simplest possible physics

\acknowledgements{We would like to thank the reviewer, M.~Pessah, for  helpful suggestions that 
led to significant improvement in the manuscript. We also thank P.~Ben\'itez-Llambay, J.~Goodman, L.~Krapp  for helpful discussions.
Support for JS  was
provided by Rutherford Discovery Fellowship RDF-U001804 and Marsden Fund grant UOO1727, which are managed through the Royal Society Te Ap\=arangi. Support for PFH was provided by NSF Collaborative Research Grants 1715847 \& 1911233, NSF CAREER grant 1455342, and NASA grants 80NSSC18K0562 and JPL 1589742.}

\paragraph*{Data availability}No new data were generated or analysed in support of this research.

%\bibliographystyle{mn2e}
%\bibliography{fullbib_formatted}

%%%%%%%%%%%%%%%%%%%%%%%%%%%%%%%%%
%%%%%%%%%%%%%%%%%%%%%%%%%%%%%%%%%
%%%%%%%%%%%%%%%%%%%%%%%%%%%%%%%%%
%%%%%%%%%%%%%%%%%%%%%%%%%%%%%%%%%

\end{document}